\documentclass
[superscriptaddress,secnumarabic,amssymb,amsmath,nobibnotes,aps,prd,nofootinbib,nopacsnumber,onecolumn,12pt]{revtex4}%
\usepackage{placeins}
\usepackage{graphicx}
\usepackage{epsf}
\usepackage{bm}
\usepackage{amsmath}
\usepackage{amsfonts}
\usepackage{amssymb}%
\providecommand{\U}[1]{\protect\rule{.1in}{.1in}}

\newcommand{\be}{\begin{equation}}
\newcommand{\ee}{\end{equation}}

\newcommand{\mincir}{\raise
-3.truept\hbox{\rlap{\hbox{$\sim$}}\raise4.truept\hbox{$<$}\ }}
\newcommand{\magcir}{\raise
-3.truept\hbox{\rlap{\hbox{$\sim$}}\raise4.truept\hbox{$>$}\ }}

\RequirePackage{latexsym}
\usepackage{hyperref}
\hypersetup{
    colorlinks=true,
    linkcolor=red,
    citecolor=blue,
    filecolor=magenta,      
    urlcolor=cyan,
    pdftitle={Averaging generalized scalar field cosmologies IV: locally rotationally symmetric Bianchi V model},
    pdfpagemode=FullScreen}  
\newcommand{\bea}{\begin{eqnarray}}
\newcommand{\eea}{\end{eqnarray}}

\def\y{\varphi}
\def\z{\zeta}

\newcommand{\R}{\mathbb{R}}

\newcommand{\bef}{\begin{figure}}  \newcommand{\eef}{\end{figure}}
\newcommand{\bec}{\begin{center}}  \newcommand{\eec}{\end{center}}

\usepackage{url} 
\usepackage{pdflscape} 
\usepackage{amsmath}
\usepackage{amssymb} 
\usepackage{epsfig} 
\usepackage{amsfonts}
\usepackage{graphicx}
\usepackage{subfigure}
\usepackage{graphicx}
\usepackage{times,fancyhdr}
\usepackage{enumitem}
\setlist[enumerate,2]{label=\roman*)}

\def\case#1/#2{\textstyle\frac{#1}{#2}}

\newcommand{\ben}{\begin{eqnarray}}
\newcommand{\een}{\end{eqnarray}}
\usepackage{mdframed}
\usepackage{grffile}
\usepackage{amssymb}
\usepackage[T1]{fontenc}

\usepackage{amssymb,amsmath,amsfonts,amsthm,graphicx, graphics, setspace}
\usepackage{bigints}
\usepackage{anysize}
\usepackage{float}
\usepackage{fancyhdr}
\usepackage{subfigure}
\usepackage{pdflscape}
\usepackage{epsfig}
\usepackage{epsf}
\usepackage{epstopdf}
\usepackage{hyperref}

\newtheorem{thm}{Theorem}

\newtheorem{lem}[thm]{Lemma}

\providecommand{\U}[1]{\protect\rule{.1in}{.1in}}

\usepackage{tikz,xcolor,hyperref}

\definecolor{lime}{HTML}{A6CE39}
\DeclareRobustCommand{\orcidicon}{%
	\begin{tikzpicture}
	\draw[lime, fill=lime] (0,0) 
	circle [radius=0.16] 
	node[white] {{\fontfamily{qag}\selectfont \tiny ID}};
	\draw[white, fill=white] (-0.0625,0.095) 
	circle [radius=0.007];
	\end{tikzpicture}
	\hspace{-2mm}
}

\foreach \x in {A, ..., Z}{%
	\expandafter\xdef\csname orcid\x\endcsname{\noexpand\href{https://orcid.org/\csname orcidauthor\x\endcsname}{\noexpand\orcidicon}}
}

\begin{document}

\title[Averaging generalized scalar field cosmologies IV: locally rotationally symmetric Bianchi V model]{Averaging generalized scalar field cosmologies IV: locally rotationally symmetric Bianchi V  model}

\author{Alfredo D. Millano\orcidA{}}
\email{alfredo.millano@alumnos.ucn.cl}
\address{Departamento  de  Matem\'aticas,  Universidad  Cat\'olica  del  Norte, Avenida. Angamos  0610,  Casilla  1280  Antofagasta,  Chile}

\author{Genly Leon\orcidB{}}
\email{genly.leon@ucn.cl}
\address{Departamento  de  Matem\'aticas,  Universidad  Cat\'olica  del  Norte, Avenida. Angamos  0610,  Casilla  1280  Antofagasta,  Chile}
\address{Institute of Systems Science, Durban University of Technology, PO Box 1334,
Durban 4000, South Africa}

\begin{abstract}

This research focuses on scalar field cosmologies with a generalized harmonic potential. Our attention is centred on the anisotropic LRS Bianchi I and III metrics, Bianchi V metrics, and their isotropic limits. We provide a comprehensive overview of the first two metrics classes and offer new findings for Bianchi V metrics. We show that the Hubble parameter is a time-dependent perturbation parameter that controls the magnitude of the error between full-system and time-averaged solutions as it decreases, such that those complete and time-averaged systems have the same asymptotic behaviour. Therefore, oscillations entering the system can be controlled and smoothed out, which simplifies the problem at hand.

\end{abstract}


\maketitle

\tableofcontents

\section{Introduction}

The cosmological paradigm is based on the assumption that the observable universe is homogeneous and isotropic. Most studies use the flat Friedmann-Lemaître-Robertson-Walker (FLRW) model. Researchers then study the evolution of perturbations, particularly within the inflationary theory. However, while inflation is the most successful explanation for the observed homogeneity and isotropy, it does not fully solve the problem since the FLRW metrics are proposed from the outset instead of starting with an arbitrary metric. Some have tried to consider an entirely arbitrary metric, which would be both inhomogeneous and anisotropic \cite{Goldwirth:1989pr}; however, these calculations are incredibly complicated. Therefore, this paper focuses on homogeneous and anisotropic cosmology to extract analytical information.

This class of geometries exhibits exciting cosmological features in inflationary and post-inflationary epochs  \cite{Misner1973, peebles1993principles}. Nine anisotropic Bianchi models exist based on the real three-dimensional Lie algebra classification. In these spacetimes, three-dimensional hypersurfaces are defined by the orbits of three isometries. An essential characteristic of the Bianchi models is that the physical variables depend only on time, which means that the field equations are a system of ordinary differential equations \cite{mx2,mx3}. In recent years, the class of anisotropic geometries has gained much interest due to anisotropic anomalies in the Cosmic Microwave Background (CMB) and large-scale structure data. The origin of asymmetry and other measures of statistical anisotropy on the largest scales of the universe is a long-standing open question in cosmology. The "Planck Legacy" temperature anisotropy data show strong evidence of violating the Cosmological Principle in its isotropic aspect \cite{Fosalba:2020gls, LeDelliou:2020kbm}.

Isotropization is a crucial concept in cosmology, as it refers to whether the universe can be described in an isotropic manner without fine-tuning. The family of spatially homogeneous Bianchi cosmologies is an important gravitational model that includes the Mixmaster Universe as well as the isotropic FLRW spacetimes, which have been extensively studied and analyzed in various research works such as \cite{mr, Mis69,mis1,mx1}.  We find different compelling, exciting solutions in General Relativity (GR) and cosmology. Some of such solutions are the FLRW models, which follow the Cosmological Principle. FLRW spacetimes are obtained as the limit of some Bianchi models where the anisotropy approaches zero. The flat, open, and closed FLRW geometries are respectively associated with the Bianchi I, III, and IX spacetimes \cite{WE}. Another exciting solution is the Taub-Kasner and Bianchi II solutions, which are part of the Belinski-Khalatnikov-Lifshitz (BKL) singularity geometric description. The singularity is present in models such as Bianchi type VIII and IX. Mixmaster dynamics form a class of spatially homogeneous solutions that show the asymptotic behaviour near the singularity and exhibit properties similar to BKL, making them a complicated oscillatory and chaotic model. The BKL conjecture is fundamental to quantum cosmology developments and efforts to quantize gravity. 

Scaling solutions are helpful, particularly in addressing the Cosmic Coincidences problem. The de Sitter solution is related to the current accelerated expansion stage of the universe, where the scale factor increases exponentially with time. Einstein's static solution allows the transition from an expanding universe to a contracting one and vice versa. The (Dirac)-Milne solution corresponds to a universe with zero acceleration, enabling the transition from a universe with decelerated expansion to an accelerated one. The Minkowski solution represents an empty universe, useful as a local approximation of spacetime in reasonably small regions and the presence of matter, as long as it does not self-gravitate. Reviews on modified gravity, which discusses cosmological problems like inflation, bounce and late-time evolution, cosmological finite-time singularities, and cosmological perturbations are   \cite{DeFelice:2010aj,  Nojiri:2010wj, Nojiri:2017ncd, Odintsov:2018uaw}. In recent literature, the Dipole Cosmological Principle has been introduced. It suggests that the Universe is maximally Copernican and is still compatible with cosmic flow. This principle is the most symmetric approach that generalizes the FLRW ansatz in light of the emerging hints of a non-kinematic component in the CMB dipole. The Einstein equations in the Dipole Cosmological Principle lead to four ordinary differential equations, instead of the two Friedmann equations in the FLRW model. The two new functions in this principle can be seen as an anisotropic scale factor that breaks the isotropy group from $SO(3)$ to $U(1)$, and a "tilt" that captures the cosmic flow velocity. The result is an axially isotropic, tilted Bianchi V/VIIh cosmology. This paradigm allows for model building, and the dynamics of the expansion rate, anisotropic shear, and tilt in various examples \cite{Krishnan:2022qbv, Krishnan:2022uar, Ebrahimian:2023svi, Allahyari:2023kfm}, which is remarkable. In particular, the study \cite{Allahyari:2023kfm} examined the cosmic evolution of the universe from the Big Bang to the future and discussed early and late-time attractors of the system for non-interacting cosmic fluids. In \cite{Orjuela-Quintana:2021zoe}, a scalar field coupled to a vector field in an LRS Bianchi-I spacetime was studied, exhibiting an oscillatory dark energy equation of state.

Several studies, including \cite{Alho:2015cza, Alho:2019pku, Fajman:2020yjb, Fajman:2021cli, Leon:2019iwj, Leon:2020ovw, Leon:2020pvt, Leon:2021lct, Leon:2021rcx, Leon:2021hxc, Leon:2020pfy} have applied averaging methods to analyze single field scalar field cosmologies, and scalar field cosmologies with two scalar fields that interact gravitationally with the matter in \cite{Chakraborty:2021vcr}. In \cite{Leon:2020pvt}, scalar field cosmology with a generalized harmonic potential was examined in flat and negatively curved FLRW and Bianchi I metrics. The conservation equations were considered with an interaction between the scalar field and matter. These references utilized asymptotic methods and the theory of averaging in nonlinear dynamical systems to obtain relevant information about the solution space.
The standard dynamical systems approach faces challenges due to oscillations that enter the nonlinear system through the Klein-Gordon (KG) equations. Therefore, analyzing the oscillations using the averaging theory in nonlinear dynamical systems is necessary. This method proves that time-dependent systems and their corresponding time-averaged versions have the same late-time dynamics. Thus, the most straightforward time-averaged system determines the future asymptotic behaviour, and late-time attractors of physical interest can be found depending on the values of free parameters.
The Hubble parameter is a time-dependent perturbation parameter that controls the magnitude of the error between full-system and time-averaged solutions as it decreases. Therefore, the oscillations entering the system through the KG equation can be controlled and smoothed out, simplifying the problem. These results suggest that the oscillations arising from harmonic functions can be "averaged out".

The dynamical systems in the cosmology program consists of the following procedures:
Utilize dynamical systems theory to accurately determine the feasible asymptotic states of cosmological models, mainly when the governing equations constitute finite systems of autonomous ordinary differential equations.
Emphasize exploring cosmological models as dynamic systems, focusing on their applications in the early Universe.
Thoroughly examine the asymptotic properties of spatially homogeneous and inhomogeneous models in general relativity.
Conduct a detailed analysis of the outcomes of scalar field models with exponential potential, both with and without barotropic matter.
Conduct a comprehensive scrutiny of the dynamic properties of cosmological models derived from effective actions. Employ asymptotic methods and averaging theory to derive relevant information about the solution space of scalar-field cosmologies.
Use computational tools to solve problems in a timely and efficient manner.

Based on references 
\cite{Leon:2020pfy,Leon:2019iwj,Leon:2020ovw,Leon:2020pvt} we started the   ``Averaging generalized scalar-field cosmologies'' program \cite{Leon:2021lct}. Asymptotic methods and averaging theory are used to investigate the solutions space of scalar-field cosmologies with non-minimal interaction with a matter source with energy density $\rho_m$ and pressure $p_m$ with a barotropic  EoS  $p_m=(\gamma-1)\rho_m$ (with barotropic index $\gamma\in[0,2]$). The scalar field oscillates in a generalized harmonic potential. We studied three cases of study: (I)  Bianchi III and open FLRW model \cite{Leon:2021lct}, (II)  Bianchi I and flat  FLRW model \cite{Leon:2021rcx}, and (III) KS and closed FLRW \cite{Leon:2021hxc}. In reference \cite{Leon:2021lct}, it was proved that for LRS Bianchi III, the late-time attractors are a  matter-dominated flat FLRW universe if  $0\leq \gamma \leq \frac{2}{3}$, a matter-curvature scaling solution if  $\frac{2}{3}<\gamma <1$, and Bianchi III flat space-time for $1\leq \gamma\leq 2$. In the first case, the matter mimics de Sitter, quintessence or zero acceleration solutions. For FLRW metric with $k=-1$, the late-time attractors are flat matter-dominated FLRW universe  if  $0\leq \gamma \leq \frac{2}{3}$ and Milne solution if $\frac{2}{3}<\gamma <2$.  In all metrics, matter-dominated flat  FLRW universe represents quintessence fluid if $0< \gamma < \frac{2}{3}$.
In reference \cite{Leon:2021rcx} for flat FLRW and LRS Bianchi I metrics, it was obtained that late-time attractors of complete and time-averaged systems are given by flat matter-dominated  FLRW solution and Einstein-de Sitter solution. Interestingly, for FLRW with negative or zero curvature and Bianchi I metric, when the matter fluid corresponds to a Cosmological Constant,  $H$ asymptotically tends to constant values depending on initial conditions. That is consistent with de Sitter's expansion. 
In addition, for FLRW models with negative curvature for any $\gamma<\frac{2}{3}$ and $\Omega_K>0$, $\Omega_K \rightarrow 0$, or when $\gamma>\frac{2}{3}$ and $\Omega_K>0$ the Universe becomes curvature dominated  ($\Omega_K \rightarrow 1$). 
For flat FLRW and dust background from the qualitative analysis performed in paper, \cite{Leon:2021rcx}, we have that
$\lim_{\tau\rightarrow +\infty}\overline{\Omega}(\tau)= \text{const.}$,  and $\lim_{\tau\rightarrow +\infty}H(\tau)=0$. Also, it was numerically proved that as $H\rightarrow 0$, the values of $\overline{\Omega}$ give an upper bound for the values $\Omega$ of the original system. Therefore, the error of the time-averaged higher-order system can also be controlled by controlling the error of the original system.  
\noindent Finally, in the KS metric, it can be proved that the global late-time attractors of complete and time-averaged systems are two anisotropic contracting solutions. These are a non-flat LRS Kasner and a  Taub (flat LRS Kasner) for $0\leq \gamma<2$, and a  matter-dominated flat FLRW universe if  $0\leq \gamma \leq \frac{2}{3}$ (mimicking de Sitter, quintessence or zero acceleration solutions). For closed cosmologies (Kantowski-Sachs, positively curved FLRW, etcetera),  the quantity $D=\sqrt{H^2 + \frac{{}^{(3)} R}{6}}$, 
where ${}^{(3)} R$ is the 3-Ricci curvature of spatial surfaces (if the congruence $\mathbf{u}$ is irrotational), plays the role of a time-dependent perturbation function that controls the magnitude of the error between the solutions of complete and time-averaged problems. For FLRW metric with $k=+1$ global late-time attractors of the time-averaged system are a flat matter-dominated contracting solution that is a sink for $1<\gamma\leq 2$,     a  matter-dominated flat FLRW universe mimicking de Sitter, quintessence or zero
acceleration solutions if  $0\leq \gamma \leq \frac{2}{3}$, and an Einstein-de Sitter solution for $0\leq\gamma<1$ and large $t$. 
However, when $D$ becomes infinite and for $\gamma\geq 1$, solutions of the complete system depart from solutions of the averaged system as $D$ is large. Then,  different from KS, for the entire system and given $\gamma>1$, the orbits do not follow the average system track as $D\rightarrow \infty$. That is somewhat different behaviour from the time-averaged system, where they are saddle.  We have shown that asymptotic methods and averaging theory are powerful tools to investigate scalar-field cosmologies with generalized harmonic potential. Therefore, these analyses complete the characterization of the whole class of homogeneous but anisotropic solutions and their isotropic limits, except LRS Bianchi V. 
This paper focuses on scalar field cosmologies with generalized harmonic potential for Bianchi V models.

The paper is structured as follows: Section \ref{Sect:2} summarizes the averaging techniques used for examining nonlinear time-dependent dynamical systems.  The attention in this section is centred on the anisotropic LRS Bianchi I and III metrics. The research also explores the homogeneous and isotropic FLRW metrics. We provide a comprehensive overview of these metrics classes. For a physical application in section \ref{sec3}, averaging techniques are used to find the late time and early time behaviour of the evolution of cosmological perturbations in vacuum for a flat FLRW metric, following the line of Ref. \cite{Alho:2020cdg}. In section 
\ref{Sect:3.4}, we discuss the Bianchi V metric, offering new qualitative findings analogous to the studies in references \cite{Leon:2021lct, Leon:2021rcx}. Section \ref{conclusions} is devoted to conclusions. 

\section{Averaging of nonlinear dynamical systems}
\label{Sect:2}
In general, in the theory of averaging for nonlinear dynamical systems, initial value problems 
\begin{align}
\dot{\mathbf x} = \mathbf{f}(\mathbf x, t, \varepsilon), \quad \mathbf x(0)=\mathbf{x}_0,
\end{align} are studied. Here, $\mathbf x$ and $\mathbf f(\mathbf x, t, \varepsilon)$ belong to $\R^n$, and $\varepsilon$ is a typically small perturbation parameter. An approach often used is to perform a Taylor expansion of $\mathbf f$ in $\varepsilon$ around $\varepsilon=0$. For the periodic averaging method, the zeroth order term usually vanishes, and the standard form of the problem becomes 
\begin{align}\label{E: standard form}
\dot{\mathbf x} = \varepsilon\,\mathbf f^1(\mathbf x, t) + \varepsilon^2\,\mathbf f^{[2]}(\mathbf x,t,\varepsilon),
\quad
\mathbf x(0) = \mathbf{x}_0,
\end{align} where $\mathbf f^1$ and $\mathbf f^{[2]}$ are $T$-periodic in $t$. 
The exponents represent the respective perturbative order, and the square bracket denotes the remainder of the series (Notation~1.5.2, p~13, \cite{SandersEtAl2010}). To approximate this problem, one can solve it for $\varepsilon=0$ (unperturbed problem) and use this solution to formulate variational equations in standard form, which can then be averaged.

To first order, the theory is then concerned with the question of to what degree solutions of~\eqref{E: standard form} can be approximated by the solutions of an associated \emph{averaged system}
\begin{align}\label{E: averaged system}
\dot{\mathbf{y}} &= \varepsilon\,\overline{\mathbf f}^1(\mathbf{y}), \quad \mathbf{y}(0)=\mathbf{x}_0,
\end{align}
with
\begin{align}\label{E: f bar}
\overline{\mathbf f}^1(\mathbf{y}) &= \frac{1}{T}\int_0^T\mathbf f^1(\mathbf{y}, s)\,\mathrm ds.
\end{align}
Using methods from the theory of averaging nonlinear dynamical systems, we can prove that time-dependent systems and their corresponding time-averaged versions exhibit the same late-time dynamics. That means that simple time-averaged methods can determine future asymptotic behaviour. Therefore, we rely on amplitude-phase variables (as defined in chapter 11 of \cite{Verhulst}, and p~ 22, 24-27, 42, 54, 361 of \cite{SandersEtAl2010}) for our analysis: 
\begin{equation}\label{E: amplitude-phase}
\dot{\phi}(t)= r(t) \cos (t-\Phi(t)), \quad \phi(t)  = r(t) \sin (t-\Phi(t)),
\end{equation}
such that
\begin{equation}\label{eq_64}
r=\sqrt{\dot \phi(t)^2+{\phi}(t)^2}, \quad \Phi =t-\tan ^{-1}\left(\frac{\phi
   (t)}{\dot \phi(t)}\right). 
\end{equation}

A recent publication examined the LRS Bianchi III  metric {\cite{Fajman:2020yjb}}
\begin{align}\label{metricLRSBIII}
\mathbf g &= -\mathrm dt^2 +a(t)^2\,\mathrm dr^2 +b(t)^2\,\mathbf g_{H^2},
\end{align}
where $\mathbf{g}_{H^2}=   d \varphi^2 +  \sinh^2 (\varphi)d \zeta^2$ denotes the 2-metric of negative constant curvature on hyperbolic 2-space.   This system can be compared to a harmonic oscillator with nonlinear damping, with the time-dependent aspect being determined by the coupling of the Einstein equations with the Klein-Gordon equation, as seen in  
\begin{equation}
\ddot \phi + \phi = H[-3\dot\phi]. \label{E: KG}
\end{equation} 

The study introduced the state vector $\mathbf x=[\Sigma, \Omega, \Phi]^{\mathrm T}$, where $\Sigma$ is a dimensionless measure of the anisotropy, $\Omega=r^2/(6H^2)$ and $(r, \Phi)$ are defined by \eqref{eq_64}. The system can be represented in quasi-standard form, as shown in Equation \eqref{E: quasi-standard form}
\begin{align}\label{E: quasi-standard form}
\begin{bmatrix} \dot H \\ \dot{\mathbf x} \end{bmatrix} &=
H\,\mathbf F^1(\mathbf x, t) + H^2\,\mathbf F^{[2]}(\mathbf x, t) =
H \begin{bmatrix} 0 \\ \mathbf f^1(\mathbf x, t) \end{bmatrix} + H^2 \begin{bmatrix} f^{[2]}(\mathbf x, t) \\ \mathbf 0 \end{bmatrix},
\end{align} which resembles the standard form \eqref{E: standard form} with $H(t)$ acting as the perturbation parameter $\varepsilon$.  
In a study by \cite{Fajman:2021cli}, averaging tools were used to analyze a system. The solution of the corresponding averaged system is referred to as $\overline{\mathbf{y}}(t)$. The study discovered that on time scales of $\mathcal O(H_*^{-1})$, where $H_*$ is the value of $H$ at a large truncation time $t^*$, $H(t^*)$, $\mathbf y(t) - \overline{\mathbf{y}}(t) = \mathcal O(H_*)$. Additionally, the study found a case of averaging with attraction, and the error estimate for the $\mathbf{x}$-components holds for all times. In a more recent study conducted by \cite{Fajman:2020yjb}, a general class of systems in the standard form \eqref{E: quasi-standard form} with $H>0$ that is strictly decreasing in $t$ and $\lim_{t\rightarrow \infty}H(t)=0$ was studied to determine the long-term behaviour of solutions.

In references \cite{Leon:2021lct,Leon:2021rcx} systems which are not in the standard form \eqref{E: quasi-standard form}, but can be expressed as a series with center in $H=0$ according to the equation
\begin{align}
\label{nonstandtard}
\begin{bmatrix}
       \dot{H} \\
        \dot{\mathbf{x}}
\end{bmatrix}= & \begin{bmatrix}
    0 \\
       \mathbf{f}^0 (\mathbf{x}, t)
 \end{bmatrix}+ H \begin{bmatrix}
       0 \\
       \mathbf{f}^1 (\mathbf{x}, t)
 \end{bmatrix}  + H^2 \begin{bmatrix}
       f^{[2]} (\mathbf{x}, t)  \\
       \mathbf{0}
\end{bmatrix}+ \mathcal{O}(H^3),
  \end{align}
  were studied. These systems depend on a parameter $\omega$, which is a free frequency that can be tuned to make $\mathbf{f}^0 (\mathbf{x}, t)= \mathbf{0}$. Therefore,  systems can be expressed in the standard form \eqref{E: quasi-standard form}. 

There, we studied a scalar-field cosmology with potential \begin{equation}
\label{pot_v2}
    V(\phi)=\mu ^2 \phi ^2 + f^2 \left(\omega ^2-2 \mu ^2\right) \left(1-\cos \left(\frac{\phi
   }{f}\right)\right), 
\end{equation}
by introducing an angular frequency $\omega \in\mathbb{R}$ through  conditions $b \mu ^3+2 f \mu ^2-f \omega ^2=0$ and $\omega^2-2 \mu^2>0$.  
Potential \eqref{pot_v2}  has the following generic features:
\begin{enumerate} 
    \item$V$ is a real-valued smooth function  $V\in C^{\infty} (\mathbb{R})$  with  $\lim_{\phi \rightarrow \pm \infty} V(\phi)=+\infty$. 
        \item $V$ is an even function  $V(\phi)=V(-\phi)$.
    \item  $V(\phi)$ has always a local minimum at $\phi=0$;  $V(0)=0, V'(0)=0, V''(0)= \omega^2> 0$.
    \item There is a finite number of values $\phi_c \neq 0$ satisfying \begin{small}
    $2 \mu ^2 \phi_c +f \left(\omega ^2-2 \mu ^2\right) \sin \left(\frac{\phi_c
   }{f}\right)=0$
    \end{small} which are local maximums or local minimums depending on whether  $V''(\phi_c)<0$ or $V''(\phi_c)>0$. For $\left|\phi_c\right| >\frac{f(\omega^2-2 \mu^2)}{2 \mu^2}= \phi_*$ this set is empty. 
    \item There exist 
    $V_{\max}= \max_{\phi\in [-\phi_*,\phi_*]} V(\phi)$   and $V_{\min}= \min_{\phi\in [-\phi_*,\phi_*]} V(\phi)=0$. The function $V$ has no upper bound but a lower bound equal to zero.
  \end{enumerate}    
Near global minimum $\phi=0$, we have 
$V(\phi) \sim \frac{\omega ^2 \phi ^2}{2}+\mathcal{O}\left(\phi ^3\right), \quad \text{as} \; \phi\rightarrow 0$. 
That is, $\omega^2$ can be related to the mass of the scalar field near its global minimum. As $\phi\rightarrow \pm \infty$  cosine- correction is bounded, then, $V(\phi) \sim \mu ^2 \phi ^2+\mathcal{O}\left(1\right) \quad \text{as} \; \phi\rightarrow \pm \infty$.  That makes it suitable to describe oscillatory behaviour in cosmology. The generalized scalar-field cosmologies with matter in LRS Bianchi III and the open FLRW model were investigated in Ref. \cite{Leon:2021lct}.

By defining $\mathbf{x}= \left(\Omega, \Sigma, \Omega_k, \Phi \right)^T$, where 
    \begin{align}
& \Omega=\sqrt{\frac{\omega^2 \phi^2+{\dot\phi}^2}{6 H^2}}, \;   \Sigma=\frac{{\sigma_+}}{H},  \;  \Omega_k=\frac{K}{3 H^2}, \;  \Phi= t \omega -\tan^{-1}\left(\frac{\omega \phi}{\dot \phi}\right),
\end{align}
and imposing the condition $b \mu ^3+2 f \mu ^2-f \omega ^2=0$,  which defines an angular frequency $\omega \in\mathbb{R}$. Then, order zero terms 
in the series expansion around $H=0$ are eliminated  assuming $\omega ^2>2 \mu ^2$ and setting $f=\frac{b \mu ^3}{\omega ^2-2 \mu ^2}$, which is equivalent to tuning $\omega$.

Hence, we obtain the following system: 
\begin{small}
\begin{subequations}
\begin{align}
 & \dot{H}= -\frac{3}{2} H^2 \Big(\gamma(1- \Sigma ^2- {\Omega_k}-\Omega^2)   +2 \Sigma^2 +2/3{\Omega_k}  + 2 \Omega
   ^2 \cos^2(t \omega -\Phi) \Big) +\mathcal{O}(H^3), \label{EQ:81b}\\
   & \dot{\mathbf{x}}= H \mathbf{f}(\mathbf{x}, t)+ \mathcal{O}(H^2), \; \mathbf{x}= \left(\Omega, \Sigma, \Omega_k, \Phi \right)^T\label{equx2}
\\
\label{EQ:108}
 & \mathbf{f}(\mathbf{x}, t) =
\begin{bmatrix}
\frac{1}{2}  \Omega  \Big(-3 (\gamma -2) \Sigma ^2+(2-3 \gamma )  {\Omega_k}   +3 \left(\Omega ^2-1\right) (-\gamma +2 \cos^2(t \omega -\Phi))\Big) \\
\frac{1}{2} \Bigg( {\Omega_k} ((2-3 \gamma ) \Sigma
   +2)   +3 \Sigma  \Big(-(\gamma -2) \Sigma ^2+\gamma   +\Omega ^2 (-\gamma +2 \cos^2(t \omega -\Phi))-2\Big)\Bigg)\\
   {\Omega_k} \Big(-3 \gamma  \left(\Sigma ^2+\Omega ^2+ {\Omega_k}-1\right)  
    +6 \Sigma ^2-2
   \Sigma + 6 \Omega ^2 \cos^2(t \omega -\Phi)  +2  {\Omega_k}-2\Big) \\
-\frac{3}{2} \sin (2 t \omega -2\Phi)
\end{bmatrix}. 
\end{align}
\end{subequations}
 \end{small}

Replacing $\dot{\mathbf{x}}= H \mathbf{f}(\mathbf{x}, t)$ where $\mathbf{f}(\mathbf{x}, t)$ is defined by \eqref{EQ:108} with $\dot{\mathbf{y}}= H  \overline{\mathbf{f}}(\mathbf{y})$, $\mathbf{y}= \left(\overline{\Omega}, \overline{\Sigma}, \overline{\Omega}_k, \overline{\Phi} \right)^T$ and $\overline{\mathbf{f}}(\mathbf{y})$ given by time-averaging    \begin{equation}
\label{timeavrg}
      \overline{\mathbf{f}}(\cdot):=\frac{1}{L} \int_{0}^L \mathbf{f}(\cdot, t) dt, \quad L=\frac{2 \pi}{\omega},  
\end{equation}
and introducing the new time variable $f^{\prime}= df/d\tau\equiv \frac{1}{H} \dot{f}$,
we obtain the averaged system: 
\begin{small}
 \begin{align}
 \label{IIIeq28}
 \begin{bmatrix}
 {H}^{\prime}\\
 {\overline{\Omega}}^{\prime}\\
 {\overline{\Sigma}}^{\prime}\\
 {{\overline{\Omega}}}_{k}^{\prime}\\
{\overline{\Phi}}^{\prime}
 \end{bmatrix}
=   \begin{bmatrix}
-\frac{1}{2} H\Big(3 \gamma  \left(1-{\overline{\Sigma}}^2-\overline{\Omega} ^2-\overline{\Omega}_k\right)     + 6 {\overline{\Sigma}}^2+ 3 \overline{\Omega}^2+ 2 \overline{\Omega}_k\Big)  \\
\frac{1}{2} \overline{\Omega}  \Big(-3 \gamma  \left(\overline{\Sigma} ^2+\overline{\Omega} ^2+\overline{\Omega}_{k}-1\right)    +6 \overline{\Sigma} ^2+3 \overline{\Omega} ^2+2 \overline{\Omega}_{k}-3\Big) \\
\frac{1}{2}   \Bigg(\overline{\Sigma}  \Big(-3 \gamma  \left(\overline{\Sigma} ^2+\overline{\Omega}
   ^2+\overline{\Omega}_{k}-1\right)   +6 \overline{\Sigma} ^2+3 \overline{\Omega} ^2+2 \overline{\Omega}_{k}-6\Big)+2 \overline{\Omega}_{k}\Bigg)\\
 - \overline{\Omega}_{k} \Big(3 \gamma  \left(\overline{\Sigma} ^2+\overline{\Omega} ^2+\overline{\Omega}_{k}-1\right)     -6 \overline{\Sigma} ^2+2 \overline{\Sigma} -3
   \overline{\Omega} ^2-2 \overline{\Omega}_{k}+2\Big) \\
   0
\end{bmatrix}.
\end{align}
 \end{small}
As in references {\cite{Alho:2015cza,Alho:2019pku,Leon:2021lct,Leon:2021rcx}}, was proven Theorem 2, \cite{Leon:2021lct}, which states:  \emph{Let $H$, $\overline{\Omega}, \overline{\Sigma}, \overline{\Omega}_k$, and $\overline{\Phi}$ be  defined  functions that  satisfy  averaged  equations
 \eqref{IIIeq28}. Then, there exist continuously differentiable functions $g_1, g_2, g_3$ and $g_4$,  such that   $\Omega, \Sigma, \Omega_k$ and $\Phi$  are locally given  by 
 \begin{small}
\begin{align}
&  \begin{bmatrix} \Omega \\ \Sigma \\ \Omega_k \\ \Phi \end{bmatrix} = \begin{bmatrix} \Omega_{0} \\ \Sigma_{0} \\ \Omega_{k0} \\ \Phi_{0} \end{bmatrix}+ H  \begin{bmatrix}
    g_1(H , \Omega_{0}, \Sigma_{0}, \Omega_{k0}, \Phi_{0}, t)\\
    g_2(H , \Omega_{0}, \Sigma_{0}, \Omega_{k0}, \Phi_{0}, t)\\
    g_3(H , \Omega_{0}, \Sigma_{0}, \Omega_{k0}, \Phi_{0}, t)\\
    g_4(H , \Omega_{0}, \Sigma_{0}, \Omega_{k0}, \Phi_{0}, t)
\end{bmatrix}. \label{AppBIIIquasilinear211}
\end{align}
\end{small}
Then,  functions $\Omega_{0}, \Sigma_{0}, \Omega_{k0}, \Phi_0$ and averaged solution $\overline{\Omega},  \overline{\Sigma}, \overline{\Omega}_k, \overline{\Phi}$  have the same limit as $t\rightarrow \infty$. 
Setting $\Sigma=\Sigma_0=0$ derived comparable results for the negatively curved FLRW model.}

In Ref. \cite{Leon:2021lct} were found the exact solutions associated with equilibrium points of the 
averaged system  \eqref{IIIeq28}. They are summarized as follows. 

\begin{enumerate}
\item $T$ is a Taub-Kasner solution ($p_1=1, p_2= 0, p_3= 0$) with $a(t)= \frac{\left(3 H_{0} t+1\right)}{c_2}$ and   $b(t)=\sqrt{c_2}$.
        
\item $Q$ is a non-flat LRS Kasner ($p_1=-1/3, p_2= 2/3, p_3= 2/3$) Bianchi I solution with $a(t)= c_1^{-2}\left( {3 H_{0} t+1}\right)^{-1/3}$  and $b(t)={c_1^{-1}}{\left(3 H_{0} t+1\right)^{2/3}}$.

\item $D$ is a Bianchi III form of flat spacetime with  $a(t)=c_1^{-1}$ and  $b(t)=\frac{(3 H_{0} t+2)}{2 \sqrt{c_1}}$.

\item $F$ is a scalar field dominated solution which for large $t$ behaves as a de Sitter solution with  $a(t)=c_1^{-1} {t^{2/3}}$  and  $b(t)={c_2^{-1/2}}{t^{2/3}}$. 

\item $F_0$ is a Matter dominated FLRW universe with  $a(t)=\ell_{0} \left(\frac{3 \gamma  H_{0} t}{2}+1\right)^{\frac{2}{3
   \gamma }}$ and $b(t)=\ell_{0} \left(\frac{3 \gamma  H_{0} t}{2}+1\right)^{\frac{2}{3
   \gamma }}$.

\item $MC$ is a matter-curvature scaling solution with  $a(t)=\ell_{0} \left(\frac{3 \gamma  H_{0} t}{2}+1\right)^{\frac{2}{3
   \gamma }}$ and  $b(t)=\ell_{0} \left(\frac{3 \gamma  H_{0} t}{2}+1\right)^{\frac{2}{3
   \gamma }}$, 
where $c_1, c_2, \ell_0 \in \mathbb{R}^+$. 
\end{enumerate}

The results from the linear stability analysis, the Center Manifold calculations, and combined with Theorem 2, \cite{Leon:2021lct} lead Theorem 3, \cite{Leon:2021lct}, which states: 
\emph{The late-time attractors of the full system and averaged system  for the Bianchi III line element are:
\begin{enumerate}
    \item[(i)]  The matter dominated FLRW universe $F_0$    with  line element 
    \begin{small}
     \begin{align}
\label{scaling2}
   &  ds^2= - dt^2 + \ell_{0}^2 \left(\frac{3 \gamma  H_{0} t}{2}+1\right)^{\frac{4}{3
   \gamma }} \left (dr^2 + \mathbf{g}_{H^2}\right), \; 0< \gamma \leq  2/3.
\end{align}    
    \end{small}
$F_0$ represents a  quintessence fluid for $0<\gamma<2/3$ or a zero-acceleration model for $\gamma=2/3$. 
In the limit $\gamma=0$, we have the de Sitter solution. 
    \item[(ii)] The matter-curvature scaling solution $MC$ with $\overline{\Omega}_m=3(1-\gamma)$ and line element  \begin{small}
     \begin{align}
   &  ds^2= - dt^2 + \ell_{0}^2 \left(\frac{3 \gamma  H_{0} t}{2}+1\right)^{\frac{4}{3
   \gamma }} \left (dr^2 + \mathbf{g}_{H^2}\right), \; 2/3<\gamma <1.
   \end{align}    
    \end{small}
        \item[(iii)] The Bianchi III flat spacetime $D$  with metric 
    \begin{small}
    \begin{align}
   &  ds^2= - dt^2 + c_1^{-2} dr^2 +  \frac{(3
   H_{0} t+2)^2}{4 c_1} \mathbf{g}_{H^2}, \; 1\leq \gamma\leq 2. \label{metricD}
\end{align}
    \end{small}
      \end{enumerate}}

The invariant set $\Sigma=0$ of the Bianchi III corresponds to negatively curved FLRW models, or \emph{open FLRW model}, with metric
\begin{align}
\label{mOpenFLRW}
   &  ds^2= - dt^2 +a(t)^2 dr^2 
 +a(t)^2  \sinh(r)^2 \big(d\y^2 + \sin^2 \y\, d\z^2\big).
\end{align}
In this invariant set, we have the system 
\begin{subequations}
\begin{align}
& \dot{\mathbf{x}}= H \mathbf{f}(\mathbf{x}, t) + \mathcal{O}(H^2), \;   \mathbf{x}= \left(\Omega, \Omega_k, \Phi \right)^T,\\
   &  \dot{H}=-H^2\Bigg[ \frac{1}{2} \left(3
   \gamma\left(1-\Omega^2- \Omega_{k}\right)+2 \Omega_{k}\right)  +3 \Omega^2 \cos ^2(t \omega -\Phi)\Bigg]+  \mathcal{O}(H^3),
\\
    \label{EQ:50}
 &  f(\mathbf{x}, t) = 
\begin{bmatrix}
   \frac{1}{2}   \Omega  \left(3 \gamma -3 \gamma  \left(\Omega^2+\Omega_{k}\right)+2 \Omega_{k}\right)  +3  \Omega \left(\Omega^2-1\right) \cos ^2(t \omega -\Phi) \\
- \Omega_{k} \left(3 \gamma  \Omega^2+(3 \gamma -2) (\Omega_{k}-1)\right)   +6  \Omega^2 \Omega_{k} \cos ^2(t \omega -\Phi)\\
    -\frac{3}{2} \sin (2 t \omega -2 \Phi)
\end{bmatrix}, 
\end{align}
\end{subequations} and the averaged system: 
\begin{align}
\label{Neq46}
\begin{bmatrix}
{\overline{\Omega}}^{\prime}\\
{\overline{\Omega}} _k^{\prime}\\
{\overline{\Phi}}^{\prime}
\end{bmatrix} 
=
\begin{bmatrix}
   -\frac{1}{2}  \; \overline{\Omega}  \left(3 (\gamma -1) \left(\overline{\Omega} ^2-1\right)+(3 \gamma -2)
   {{\overline{\Omega}_{k}}}\right)\\
   -\; {{\overline{\Omega}_{k}}} \left(3 (\gamma -1) \overline{\Omega} ^2-3 \gamma +(3 \gamma -2) {{\overline{\Omega}_{k}}}+2\right)\\
   0 
\end{bmatrix}. 
\end{align}
Exact solutions associated with equilibrium points of the 
averaged system  \eqref{Neq46} are \cite{Leon:2021lct}: 
\begin{enumerate}
    \item   $F$ is a scalar field dominated solution which for large $t$ behaves as a de Sitter solution with $a(t)= a_{0}  \left(\frac{3    H_{0} t}{2}+1\right)^{\frac{2}{3}}$. 
    \item $F_0$ is a matter dominated FLRW universe with   $a(t)=a_{0} \left(\frac{3 \gamma  H_{0} t}{2}+1\right)^{\frac{2}{3
   \gamma }}$.
    \item   $C$ is a Milne solution with   $a(t)=a_{0}  \left(H_0 t+1\right)  $, where      $a(t)$ denotes a scale factor of metric \eqref{mOpenFLRW} and $ a_0 \in \mathbb{R}^+$.
\end{enumerate}

The results from the linear stability analysis, the Center Manifold calculations, and combined with Theorem 2 (for $\Sigma=0$, open FLRW), \cite{Leon:2021lct} lead to Theorem 4, \cite{Leon:2021lct}, which states: 
\emph{The late-time attractors of  the full system   and the averaged system  are: 
\begin{enumerate}
    \item[(i)] The matter dominated FLRW universe $F_0$
   with line element        \begin{small}
    \begin{align}
\label{scaling}
   &  ds^2= - dt^2 + a_{0}^2 \left(\frac{3 \gamma  H_{0} t}{2}+1\right)^{\frac{4}{3
   \gamma }} \left (dr^2 +\sinh^2 r d\Omega^2 \right), \; 0< \gamma \leq 2/3,
\end{align}
\end{small}
\newline where
$d\Omega^2 = d\y^2 + \sin^2 \y\, d\z^2$
is the metric for a two-sphere. $F_0$ represents a  quintessence fluid or a zero-acceleration model for $\gamma=2/3$. 
In the limit $\gamma=0$, we have the de Sitter solution. 
\item[(ii)]   The Milne solution $C$ with $\overline{\Omega}_k=1, k=-1$ with  line element \begin{align}
   &  ds^2= - dt^2 + a_{0}^2 \left(H_0 t+1\right)^{2} \left( dr^2+
\sinh^2 r d\Omega^2 \right), \; 2/3<\gamma <2. \label{Milne}
\end{align}
    \end{enumerate}}

Ref. \cite{Leon:2021rcx} investigated a generalized scalar-field cosmology with the matter in LRS Bianchi I and the flat  FLRW model. 
For \emph{LRS Bianchi I} model is used, the metric 
\begin{align}
\label{metricLRSBI}
   &  ds^2= - dt^2 + a^2(t) dr^2 
 + b^2(t)  \left(d \varphi^2 +  \varphi^2 d \zeta^2\right),
\end{align} 
where the functions $a(t)$ and $b(t)$ are interpreted as the scale factors. 

Denoting $\mathbf{x}= \left(\Omega, \Sigma, \Phi \right)^T$ and  using the condition $b \mu ^3+2 f \mu ^2-f \omega ^2=0$, we obtain: 
\begin{small}
\begin{subequations}
\begin{align}
& \dot{H}= -\frac{3}{2} H^2 \Big[\gamma  \left(1-\Sigma ^2-\Omega ^2\right) +2 \Sigma ^2+ 2\Omega ^2 \cos^2(\Phi -t
   \omega )\Big]   + \mathcal{O}(H^3), \label{EQ:61b}\\
& \dot{\mathbf{x}}= H \mathbf{f}(\mathbf{x}, t)+  \mathcal{O}(H^2), \label{equx}
\\
\label{EQ:87}
   & \mathbf{f}(\mathbf{x}, t) =
   \begin{bmatrix}
\frac{3}{2} \Omega  \big(\gamma  \left(1-\Sigma ^2-\Omega ^2\right)+2 \Sigma ^2  
+2 \left(\Omega ^2-1\right) \cos^2(\Phi -t \omega )\big) \\
 \frac{3}{2} \Sigma  \big(-\gamma  \left(\Sigma ^2+\Omega
   ^2-1\right)+2 \Sigma ^2 
   +2\Omega ^2 \cos^2  (\Phi -t \omega )-2\big)\\
-\frac{3}{2} \sin (2 t \omega -2\Phi)
\end{bmatrix}.
\end{align}
\end{subequations}
\end{small}
Replacing $\dot{\mathbf{x}}= H \mathbf{f}(\mathbf{x}, t)$ with $\mathbf{f}(\mathbf{x}, t)$ as defined in \eqref{EQ:87} 
by $\dot{\mathbf{y}}= H  \overline{\mathbf{f}}(\mathbf{y})$ with  $\mathbf{y}= \left(\overline{\Omega}, \overline{\Sigma}, \overline{\Phi} \right)^T$ and $\overline{\mathbf{f}}$ as defined by \eqref{timeavrg}, and introducing the time variable $f^{\prime}\equiv \frac{1}{H} \dot{f}$,
we obtain the averaged system: 
\begin{small}
\begin{align}
\label{Ieq27}
\begin{bmatrix}
 {H}^{\prime}\\
 {\overline{\Omega}}^{\prime}\\
 {\overline{\Sigma}}^{\prime}\\
 {\overline{\Phi}}^{\prime}
\end{bmatrix}
  =\begin{bmatrix}
  -\frac{3}{2} H 
   \left(\gamma  \left(1-\overline{\Sigma}^2-\overline{\Omega}^2\right)+2 \overline{\Sigma}^2+\overline{\Omega }^2\right)\\
    \frac{3}{2} \overline{\Omega}  \left(\gamma  \left(1-\overline{\Sigma} ^2-\overline{\Omega }^2\right)+2 \overline{\Sigma} ^2+\overline{\Omega}^2-1\right)\\
    \frac{3}{2}  \overline{\Sigma}  \left(\gamma  \left(1-\overline{\Sigma} ^2-\overline{\Omega}^2\right)+2 \overline{\Sigma}^2+\overline{\Omega}^2-2\right)\\
    0
    \end{bmatrix}.
\end{align}
\end{small}
Proceeding analogously as in references \cite{Alho:2015cza, Alho:2019pku} but for three-dimensional systems  instead of a 1-dimensional one, was proved Theorem 1 of \cite{Leon:2021rcx} which states: \emph{Let  the functions $H$, 
$\overline{\Omega}, \overline{\Sigma}$, and $\overline{\Phi}$ be defined as solutions of the averaged equations  \eqref{Ieq27}. Then, there exist continuously differentiable functions $g_1, g_2$ and $g_3$ such that  $\Omega, \Sigma, \Phi$ are locally given by a nonlinear transformation
\begin{small}
\begin{align}
& \begin{bmatrix} \Omega \\ \Sigma \\ \Phi \end{bmatrix} = \begin{bmatrix} \Omega_{0} \\ \Sigma_{0}  \\ \Phi_{0} \end{bmatrix} + H  \begin{bmatrix}
    g_1(H , \Omega_{0}, \Sigma_{0}, \Phi_{0}, t)\\
    g_2(H , \Omega_{0}, \Sigma_{0}, \Phi_{0}, t)\\
    g_3(H , \Omega_{0}, \Sigma_{0}, \Phi_{0}, t)
 \end{bmatrix}, 
\end{align}
\end{small}
where $\Omega_{0}, \Sigma_{0}, \Phi_{0}$ are zero order approximations of $\Omega, \Sigma, \Phi$ as $H\rightarrow 0$. Then, functions $\Omega_{0}, \Sigma_{0}, \Phi_0$ and averaged solution $\overline{\Omega},  \overline{\Sigma}, \overline{\Phi}$  have the same limit as $t\rightarrow \infty$.
Setting $\Sigma=\Sigma_0=0$ matching results for flat FLRW model are derived.}

Exact solutions associated with equilibrium points of the 
averaged system \eqref{Ieq27} are \cite{Leon:2021rcx}: 
\begin{enumerate}
 \item $T$ is a Taub-Kasner solution ($p_1=1, p_2= 0, p_3= 0$) with $a(t)=\frac{\left(3 H_{0} t+1\right)}{c_2}$  and $b(t)=\sqrt{c_2}$ where $c_1, c_2, \ell_0 \in \mathbb{R}^+$. 

\item $Q$ is a non-flat LRS Kasner ($p_1=-1/3, p_2= 2/3, p_3= 2/3$) Bianchi I solution with $a(t)=c_1^{-2}\left( {3 H_{0} t+1}\right)^{-1/3}$ and  $b(t)={c_1^{-1}}{\left(3 H_{0} t+1\right)^{2/3}}$.

\item $F$ is a scalar field dominated solution with $a(t)=c_1^{-1} {t^{2/3}}$  and $b(t)={c_2^{-1/2}}{t^{2/3}}$ which for large $t$ behaves as a de Sitter solution.

 \item        $F_0$ is a Flat matter dominated FLRW universe with $a(t)=b(t)=a_{0} \left(\frac{3 \gamma  H_{0} t}{2}+1\right)^{\frac{2}{3
   \gamma }}$. 
     \end{enumerate}
 Results from the linear stability analysis, which are combined with Theorem 1 of \cite{Leon:2021rcx}, lead to the following Theorem 2, \cite{Leon:2021rcx}: 
 \emph{The late-time attractors of the full system and time-averaged system  for the LRS Bianchi I line element are:
\begin{enumerate}
    \item[(i)]  The flat matter dominated FLRW Universe $F_0$  with  the line element  \begin{align}
   &  ds^2= - dt^2 + \ell_{0}^2 \left(\frac{3 \gamma  H_{0} t}{2}+1\right)^{\frac{4}{3
   \gamma }} \left[dr^2  +   d \varphi^2 +  \varphi^2 d \zeta^2 \right], \; 0< \gamma < 1. \label{eq60}
\end{align}
$F_0$ represents a  quintessence fluid if $0<\gamma<2/3$ or a zero-acceleration model if $\gamma=2/3$. 
Taking limit $\gamma=0$ we have $\ell(t)=\ell_{0} \left(\frac{3 \gamma  H_{0} t}{2}+1\right)^{\frac{2}{3
   \gamma }}\rightarrow \ell_0  e^{H_0 t}$, i.e., a de Sitter solution.
  \item[(ii)] The scalar field dominated solution $F$   with  line element \begin{align}
   &  ds^2= - dt^2 + c_1^{-2} {t^{4/3}} dr^2  +  {c_2^{-1}}{t^{4/3}}  \left[ d \varphi^2 +  \varphi^2 d \zeta^2\right], \;1<\gamma\leq 2. \label{eq61}
\end{align}
For large $t$, the equilibrium point can be associated with the de Sitter solution.
\end{enumerate}}
For $\gamma=1$, $F_0$ and $F$ are stable because they belong to the stable normally hyperbolic line of equilibrium points. For $\gamma=2$, $F$ is asymptotically stable.

\emph{Flat FLRW metric.} The metric is given by \begin{align}
\label{metricFLRW}
& ds^2 = - dt^2 + a^2(t) \Big[ dr^2+  r^2 (d\y^2 + \sin^2 \y\, d\z^2)\Big],
\end{align}
In this case, the field equations are obtained  by setting $k=0$ and by substituting $\Sigma=0$ to obtain the Taylor expansion:
\begin{subequations}
\begin{align}
& \dot{H}= -H^2 \left(\frac{3}{2} \gamma\left(1-\Omega^2\right)  +3 \Omega^2 \cos ^2(t \omega -\Phi)\right)+  \mathcal{O}(H^3), \\
& \dot{\mathbf{x}}= H \mathbf{f}(\mathbf{x}, t) + \mathcal{O}(H^2), \;   \mathbf{x}= \left(\Omega,  \Phi \right)^T, 
\\
\label{2EQ:50}
 &  \mathbf{f}(\mathbf{x}, t)  = 
   \begin{bmatrix}
   \frac{3}{2} \gamma\left(1-\Omega^2\right) +3  \Omega \left(\Omega^2-1\right) \cos ^2(t \omega -\Phi) \\
    -\frac{3}{2} \sin (2 t \omega -2 \Phi)
\end{bmatrix}. 
\end{align}
\end{subequations}
Replacing $\dot{\mathbf{x}}= H \mathbf{f}(\mathbf{x}, t)$ and   $\mathbf{f}(\mathbf{x}, t)$ as defined by \eqref{2EQ:50} with $\dot{\mathbf{y}}= H  \overline{f}(\mathbf{y})$  where  $\mathbf{y}= \left(\overline{\Omega},  \overline{\Phi} \right)^T$  with the time averaging \eqref{timeavrg}, and introducing the new time variable $f^{\prime}\equiv \frac{1}{H} \dot{f}$, we obtain the following time-averaged system: 
\begin{align} 
\label{eq46}
\begin{bmatrix}
{\overline{\Omega}}^{\prime}   \\ 
{\overline{\Phi}}^{\prime}
\end{bmatrix} =\begin{bmatrix}
-\frac{3}{2} \; \overline{\Omega}   (\gamma -1) \left(\overline{\Omega} ^2-1\right)\\
0
\end{bmatrix}.
\end{align}
The time-averaged Raychaudhuri equation for flat FLRW metric is obtained by setting $\overline{\Sigma}=0$ in eq. \eqref{Ieq27}, which corresponds to flat FLRW models.

Exact solutions associated with equilibrium points of the 
averaged equations \eqref{eq46} are \cite{Leon:2021rcx}: 
\begin{enumerate}
\item $F$ with $a(t)= a_{0}  \left(\frac{3    H_{0} t}{2}+1\right)^{\frac{2}{3}}$
  where $a_0 \in \mathbb{R}^+$. 
  
\item $F_0$ is a flat matter dominated FLRW universe with $a(t)=a_{0} \left(\frac{3 \gamma  H_{0} t}{2}+1\right)^{\frac{2}{3\gamma }}$.
\end{enumerate}
Results from the linear stability analysis, which are combined with Theorem 1 of \cite{Leon:2021rcx} (for $\Sigma=0$) and lead to Theorem 3, \cite{Leon:2021rcx}, which states:
\emph{The late-time attractors of the full system  and  averaged system  with $\Sigma=0$   are: 
\begin{enumerate}
    \item[(i)] The flat matter dominated FLRW Universe $F_0$
   with line element      \begin{align}
   &  ds^2= - dt^2 + a_{0}^2 \left(\frac{3 \gamma  H_{0} t}{2}+1\right)^{\frac{4}{3
   \gamma }} \left[ dr^2  + r^2  \left(d \varphi^2 + \varphi^2 d \zeta^2\right)\right],\;0\leq \gamma < 1.  \label{eq69}
\end{align} 
$F_0$ represents a  quintessence fluid if $1<\gamma<2/3$ or a zero-acceleration model if $\gamma=2/3$. 
We have $a(t)=a_{0} \left(\frac{3 \gamma  H_{0} t}{2}+1\right)^{\frac{2}{3
   \gamma }}\rightarrow a_0  e^{H_0 t}$ as $\gamma\rightarrow 0$, i.e.,  a de Sitter solution is recovered.
\item[(ii)]    The scalar field dominated solution $F$   with  line element    \begin{align}
   &  ds^2= - dt^2 + a_{0}^2 \left(\frac{3    H_{0} t}{2}+1\right)^{\frac{4}{3
     }} \left[ dr^2 + r^2  \left(d \varphi^2 + \varphi^2 d \zeta^2\right)\right], \; 1<\gamma\leq 2  \label{eq70}
\end{align}    
For large $t$, the equilibrium point can be associated with the de Sitter solution.
    \end{enumerate}}

\section{Evolution of Cosmological perturbations in vacuum}\label{sec3}

In this section, following the line of Ref. \cite{Alho:2020cdg}, we investigate the dynamics of linear scalar cosmological perturbations for a generic scalar field model by the methods of dynamical systems. We use the perturbation of a scalar field $\phi_0$ in the background. The most generic scalar perturbed FLRW metric can be written as \cite{Bardeen:1980kt}
\begin{small}
\begin{equation}
    ds^2 = - \left(1+\alpha\right)dt^2 - 2a(t)\left(\beta,_{i}-S_{i}\right)dtdx^i + a^{2}(t)\left[\left(1+2\psi\right)\delta_{ij} + 2\partial_{i}\partial_{j}\gamma + 2\partial_{(i}f_{j)} + h_{ij}\right]dx^{i}dx^{j},
\end{equation}
\end{small}
where the inhomogeneous perturbation quantities $\alpha,\,\beta,\,\psi,\,\gamma,\,f_i,\,h_{ij}$ are functions of both $t$ and $\vec{x}$. The quantity $\psi(t,\vec{x})$ is directly related to the 3-curvature of the spatial hyper-surface
\begin{equation}
    ^{(3)}R = - \frac{4}{a^2}\nabla^{2}\psi.
\end{equation}
For a scalar field, one also needs to take into account the perturbation of the scalar field $\delta\phi(t,\vec{x})$ and, for a perfect fluid, the perturbed energy-momentum tensor is
\begin{equation}
    T^0_0 = -\left(\rho(t) + \delta\rho(t,\vec{x})\right), \quad
    T^0_i = -\left(\rho(t) + P(t)\right)\partial_{i}v(t,\vec{x}), \quad
    T^i_j = \left(P(t) + \delta P(t,\vec{x})\right)\delta^i_j,
\end{equation}
being $v(t,\vec{x})$ the velocity potential. In what follows, we will restrict ourselves to the case when there is no matter, but only a scalar field is present. The reason is simplicity. 
Suppose one wants to investigate cosmological perturbations in the presence of two matter components, e.g. a perfect fluid and a scalar field. In that case, one needs to consider entropy perturbations as well. A widespread practice in literature concentrates on a particular cosmological epoch when only one matter component is dominant. In that sense, even though not generic, our subsequent analysis is still relevant when the Universe is a scalar field dominated, e.g. during the early inflationary epoch or the late-time acceleration. 
Of course, there is the gauge issue; the perturbation quantities defined above are not gauged invariant \cite{Mukhanov:1990me, Brandenberger:1992dw, Brandenberger:1993zc, Brandenberger:1992qj, DeFelice:2010aj}. In this sense, various gauge-invariant perturbation quantities have been introduced in the literature. In this article, we will consider the following three gauge-invariant perturbation quantities.
There are several approaches for cosmological perturbations. The three more relevant are the Bardeen potentials introduced by James Bardeen in \cite{Bardeen:1980kt}, who gave the first-ever gauge-invariant formulation for cosmological perturbations. These quantities are gauge-invariant perturbation quantities constructed solely out of metric perturbations. There are two such quantities in which, for the case of a single scalar field, both the Bardeen potentials are equal, and they were examined in \cite{Bardeen:1980kt, Mukhanov:1990me, Brandenberger:1992dw, Brandenberger:1993zc, Brandenberger:1992qj}. 
Moreover, for single scalar field models, we can define comoving curvature perturbation coinciding with the 3-curvature perturbation of the spatial slice in the \emph{comoving} gauge, which is given by $\delta\phi=0$ for single scalar field models. The evolution of comoving curvature perturbation at linear order was investigated in \cite{DeFelice:2010aj}. Another gauge-invariant perturbation variable that we will consider is the so-called Sasaki-Mukhanov variable \cite{Kodama:1984ziu, Mukhanov:1988jd}, or the scalar field perturbation in uniform curvature gauge, defined as
$\varphi_c \equiv \delta\phi - \frac{\dot{\phi}}{H}\psi$. At the linear level, this variable follows the perturbation equation, which is valid strictly only in the absence of matter, 
\begin{align}\label{3h}
& \frac{d^2\varphi_c}{dN^2}+\frac{d\varphi_c}{dN}\left(H^{-2} V\right) + H^{-2}\left(V_{,\phi \phi} + 2\frac{\dot{\phi}}{H} V_{,\phi}+\left(\frac{\dot{\phi}}{H}\right)^2 V\right)\varphi_c - {\mathcal{H}^{-2}}\nabla^2 \varphi_c = 0.
 \end{align}
To obtain a dynamical system that describes the evolution of perturbations, we first introduce Cartesian spatial coordinates and make the Fourier transform of the perturbation variables. This results in 
$\mathcal{H}^{-2} \nabla^2 \longrightarrow - k^2 \mathcal{H}^{-2}$.  

At the background level, we consider a scalar field in vacuum ($\rho_m=0$) and a flat FLRW metric under the potential 
\begin{equation}
    V(\phi)=  \frac{\phi^2}{2} +f\left[1- \cos \left(\frac{\phi }{f}\right)\right].
\end{equation} 
Therefore, the KG equation is 
\begin{equation}
\label{KGFLRW}
    \ddot\phi+3 H \dot \phi +\phi + \sin\left( \frac{\phi}{f}\right)= 0,
\end{equation}
and the Raychaudhuri equation is 
\begin{equation}
     \dot{H}=-\frac{1}{2} {\dot{\phi}}^2.
\end{equation}
Using the rules
\begin{small}
\begin{equation}
    \frac{d}{d t}= H \frac{d}{d\tau }, \quad   \frac{d^2}{d t^2}= H^2 \left(\frac{d^2}{d\tau ^2} -(1+ q) \frac{d}{d\tau }\right), \quad   q:= -1-\frac{\dot{H}}{H^2}=\frac{\dot{\phi}^2}{2 H^2} -1 , 
\end{equation}
\end{small}
in equation \eqref{KGFLRW}, then, we acquire the system 
\begin{small}
\begin{subequations}
\begin{align}
&\frac{d^2  \phi}{d\tau ^2} -\left(\frac{\dot{\phi}^2}{2 H^2} -3\right) \frac{d  \phi}{d\tau } + \frac{\phi + \sin\left( \frac{\phi}{f}\right)}{H^2}= 0, 
\\
& \frac{d^2\varphi_c}{dN^2}+\frac{1}{H^2}\left(\frac{\phi^2}{2} +f\left[1- \cos \left(\frac{\phi }{f}\right)\right]\right) \frac{d\varphi_c}{dN} \nonumber \\
& + \left( \frac{\frac{1+ \cos \left(\frac{\phi }{f}\right)}{f}+ 2\left(\sin \left(\frac{\phi }{f}\right)+\phi\right)\frac{\dot{\phi}}{H}+ \left(\frac{\phi^2}{2} +f\left[1- \cos \left(\frac{\phi }{f}\right)\right]\right)\left(\frac{\dot{\phi}}{H}\right)^2}{H^2}\right)\varphi_c  + \frac{k^2}{H^2 a^2} \varphi_c = 0. \label{pert}
 \end{align}
 \end{subequations}
\end{small}
In equation \eqref{pert}, we first note that $\varphi_c$ is generally complex (as it came from Fourier transformation). 
So, we write $\varphi_c=F_1+if_2$, where $F_1$ and $F_2$ are the real and imaginary parts of $\varphi_c$, respectively. Moreover, the resulting equation   has the structure 
\begin{equation}
 F{^{\prime \prime}}+P F^{\prime}+Q F =0, \label{pert-eq}
\end{equation}
where the primes denote the derivative with respect to $N$, 
\begin{small}
\begin{align}
    & P= \frac{1}{H^2}\left(\frac{\phi^2}{2} +f\left[1- \cos \left(\frac{\phi }{f}\right)\right]\right), 
    \end{align}
    \begin{align}
    & Q=  \frac{\frac{1+ \cos \left(\frac{\phi }{f}\right)}{f}+ 2\left(\sin \left(\frac{\phi }{f}\right)+\phi\right)\frac{\dot{\phi}}{H}+ \left(\frac{\phi^2}{2} +f\left[1- \cos \left(\frac{\phi }{f}\right)\right]\right)\left(\frac{\dot{\phi}}{H}\right)^2}{H^2} + \frac{k^2}{a^2H^2},
\end{align}
\end{small}
that is the same for $F_1$ and $F_2$. Generically, we denote $F_i=r_i\cos\theta_i$ and $F^{\prime}_i=r_i\sin\theta_i$ where $i=1,2$. So,
$ F^{\prime}=F  \tan\theta$, 
where $\frac{F^{\prime}}{F}=y$ has a period of $\pi$. Hence, the mapping $y=\tan\theta$ is two-to-one and, therefore, when $\theta$ makes one revolution ($0 \rightarrow  2\pi$), $y$ has to be traversed twice $-\infty \rightarrow +\infty$ \cite{Alho:2020cdg}.

Following this line, Equation \eqref{pert-eq} can be expressed then as 
\begin{equation}
{y}{^{\prime}}=-{y} ^2  -P {y} -Q,
\end{equation}
or, alternatively, 
\begin{small}
\begin{align}
& \frac{d \theta}{d\tau }  = -\sin^2\theta- \left[\frac{1}{H^2}\left(\frac{\phi^2}{2} +f\left[1- \cos \left(\frac{\phi }{f}\right)\right]\right)\right] \sin\theta \cos\theta \nonumber \\
& -\left[\frac{\frac{1+ \cos \left(\frac{\phi }{f}\right)}{f}+ 2\left(\sin \left(\frac{\phi }{f}\right)+\phi\right)\left(\frac{d \phi}{d\tau }\right)+ \left(\frac{\phi^2}{2} +f\left[1- \cos \left(\frac{\phi }{f}\right)\right]\right)\left(\frac{d \phi}{d\tau }\right)^2}{H^2} +  \frac{k^2}{a^2H^2}\right]\cos^2\theta.
\end{align}
\end{small}
We also note that it is possible to get $f_i$ from $y_i$ through the expression
\begin{align}
 F_i(\tau)=F_i(0)\exp\left(\int_0^\tau y_i(\Tilde{\tau})d\Tilde{\tau}\right).
\end{align}

The sign of $\tan \theta$ denotes whether the $F_i(\tau)$ (for $i=1$ it is the real part) will grow or decay as $\theta$ ranges from $(-\pi,\pi]$.

Defining the variable 
\begin{equation}
 z=   k^2 a^{-2}H^{-2},
\end{equation}
we deduce the equations 
\begin{subequations}
\label{perts}
\begin{align}
& \frac{d^2  \phi}{d\tau ^2} =  \left[ \frac{1}{2} \left(\frac{d \phi}{d\tau }\right)^2 -3\right] \frac{d  \phi}{d\tau } - \frac{\phi + \sin\left( \frac{\phi}{f}\right)}{H^2}, \\
&\frac{d y}{d\tau }= -{y} ^2  - \frac{1}{H^2}\left(\frac{\phi^2}{2} +f\left[1- \cos \left(\frac{\phi }{f}\right)\right]\right)  {y} \nonumber\\
& -  \frac{\frac{1+ \cos \left(\frac{\phi }{f}\right)}{f}+ 2\left(\sin \left(\frac{\phi }{f}\right)+\phi\right)\left(\frac{d \phi}{d\tau }\right)+ \left(\frac{\phi^2}{2} +f\left[1- \cos \left(\frac{\phi }{f}\right)\right]\right)\left(\frac{d \phi}{d\tau }\right)^2}{H^2} - z, 
\end{align}
\begin{align}
& \frac{d z}{d\tau } =\left[\left(\frac{d \phi}{d\tau }\right)^2 -2\right] z, 
\\
& \frac{d H}{d\tau } = -\frac{1}{2} H \left(\frac{d \phi}{d\tau }\right)^2. 
\end{align}
\end{subequations}
Deepening into the interpretation of the variable $z= \frac{k^2}{\mathcal{H}^2}$, with $\mathcal{H}= a H$. Perturbations with $k^2 \mathcal{H}^{-2}\ll 1$ are called long wavelength or super-horizon. Those with $k^2 \mathcal{H}^{-2}\gg 1$ are considered short wavelengths or sub-horizon. Long wavelength perturbations are usually studied by choosing the idealized limiting value $k=0$, corresponding to $z = 0$.
On the other hand, short wavelength perturbations correspond to $z\rightarrow \infty$. We also note that in
choosing $z= \frac{k^2}{\mathcal{H}^2}$ as a dynamical variable, we have that the wave number
$k$ is absorbed in the definition when formulating the dynamical system. However, if we choose the reference time $t=t_U$ (i.e.,
when $N:= \ln a = 0$) to be the time for setting initial data in the state space, then different choices of $z_0= \frac{k^2}{H_0^2}$ for a given $H_0$, and $a(t_U)=1$, yield solutions with different wave
number $k$ \cite{Alho:2020cdg}.

Defining $\mathbf{x}= \left(\Omega,  \Phi, y, z \right)^T$, where 
    \begin{align}
& \Omega=\sqrt{\frac{\omega^2 \phi^2+ H^2 \left(\frac{d \phi}{d\tau }\right)^2}{6 H^2}}, \quad \Phi= t \omega -\tan^{-1}\left(\frac{\omega \phi}{H} \Big/\frac{d \phi}{d\tau }\right),
\end{align}
with inverse
\begin{align}
& \phi = \frac{H}{\omega} \sqrt{6} \Omega \sin\left( \omega t - \Phi\right),  \quad  \frac{d \phi}{d\tau } = \sqrt{6} \Omega \cos\left( \omega t - \Phi\right).
\end{align}
Recalling 
\begin{equation}
    \frac{d t}{d\tau }= \frac{1}{H},
\end{equation}
we deduce the dynamical system 
\begin{small}
\begin{subequations}
\label{perts2}
\begin{align}
& \frac{d \Omega}{d\tau }= \frac{\cos (\Phi -t \omega ) \sin \left(\frac{\sqrt{6} H \Omega  \sin (\Phi -t \omega )}{f \omega }\right)}{\sqrt{6} H^2}-\frac{\left(\omega ^2-1\right)
   \Omega  \sin (2 (\Phi -t \omega ))}{2 H \omega } \nonumber \\
& +3 \Omega \left(\Omega ^2-1\right) \cos ^2(\Phi -t \omega), \\
& \frac{d \Phi}{d\tau }= -\frac{\sin (\Phi -t \omega ) \sin \left(\frac{\sqrt{6} H   \Omega  \sin (\Phi -t \omega )}{f \omega }\right)}{\sqrt{6}
   H^2 \Omega }+\frac{\left(\omega ^2-1\right) \sin ^2(\Phi -t    \omega )}{H \omega }+\frac{3}{2} \sin (2 (\Phi -t \omega    )),   
\end{align}
\begin{align}
&\frac{d y}{d\tau }= \frac{1}{H^2} \Bigg[-12 f \Omega ^2 \cos ^2(\Phi -t \omega ) \sin ^2\left(\frac{\sqrt{\frac{3}{2}} H \Omega  \sin (\Phi -t \omega )}{f \omega }\right) \nonumber \\
& -\frac{\cos \left(\frac{\sqrt{6} H \Omega  \sin (\Phi -t \omega )}{f \omega }\right)+1}{f}+2 \sqrt{6} \Omega  \cos (\Phi -t \omega ) \sin \left(\frac{\sqrt{6} H \Omega  \sin (\Phi -t \omega )}{f  \omega }\right) \nonumber \\
& -2 f y \sin ^2\left(\frac{\sqrt{\frac{3}{2}}  H \Omega  \sin (\Phi -t \omega )}{f \omega }\right)\Bigg] \nonumber \\
& +\frac{6 \Omega ^2 \sin (2 (\Phi -t \omega
   ))}{H \omega }-\frac{9 \Omega ^4 \sin ^2(2 (\Phi -t \omega
   ))}{2 \omega ^2}-\frac{3 y \Omega ^2 \sin ^2(\Phi -t \omega
   )}{\omega ^2}-y^2-z,
\\
& \frac{d z}{d\tau } =2\left[3 \Omega^2 \cos^2\left( \omega t - \Phi\right) -1\right] z, 
\\
& \frac{d H}{d\tau } = -3 H \Omega^2 \cos^2\left( \omega t - \Phi\right). 
\end{align}
\end{subequations}
\end{small}
Taking series around $H=0$ and setting  $\omega ^2>1$ and $f=(\omega^2-1)^{-1}$ we obtain the following system
\begin{small}
\begin{subequations}
\begin{align}
&\Omega^{\prime}= 3 \Omega \left(\Omega ^2-1\right) \cos ^2(\Phi -t \omega )-\frac{H \left(\omega ^2-1\right)^3 \Omega ^3 \sin ^3(\Phi -t \omega ) \cos (\Phi -t \omega )}{\omega ^3},\\
&\Phi^{\prime}=   \frac{3}{2} \sin (2 (\Phi -t \omega )) + \frac{H \left(\omega ^2-1\right)^3 \Omega ^2 \sin ^4(\Phi -t \omega )}{\omega ^3},\\
&y^{\prime}=  \frac{2-2 \omega ^2}{H^2}+\frac{6 \omega \Omega ^2 \sin (2 (\Phi -t \omega ))}{H}    \nonumber \\
& -\frac{3 \Omega ^2 \sin ^2(\Phi -t \omega ) \left(3 \omega ^2 \Omega ^2 \cos (2 (\Phi -t \omega ))+\omega ^2 \left(-\omega ^4+3 \omega ^2+y+3 \Omega ^2-3\right)+1\right)}{\omega ^2}-y^2-z \nonumber \\
& -\frac{12 H \left(\omega ^2-1\right)^3 \Omega ^4 \sin ^3(\Phi -t \omega ) \cos (\Phi -t \omega )}{\omega ^3},\\
& z^{\prime} = z \left(6 \Omega ^2 \cos ^2(\Phi -t \omega )-2\right),\\
& H^{\prime} = -3 H \Omega ^2 \cos ^2(\Phi -t \omega ).
\end{align}
\end{subequations}
\end{small}
In the equation for $y$, the leading terms as $H \rightarrow 0$ are 
\begin{equation}
 y'=\frac{2-2 \omega ^2}{H^2},\\
\end{equation}
Hence, defining 
\begin{equation}
    v=  H^2 y,
\end{equation}
we obtain 
\begin{equation}
  v^{\prime}=-6\Omega^2 \cos^2(\Phi-t \omega)v+2 -2 \omega ^2 . 
\end{equation}
\begin{figure}
    \centering
    \includegraphics[height=6cm]{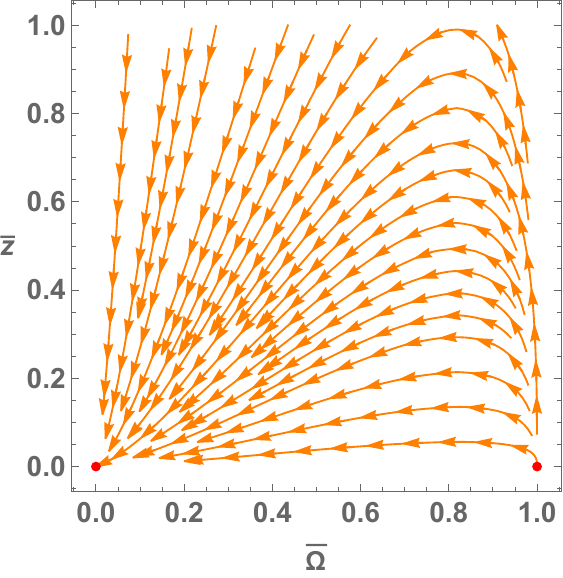}
     \includegraphics[height=6cm]{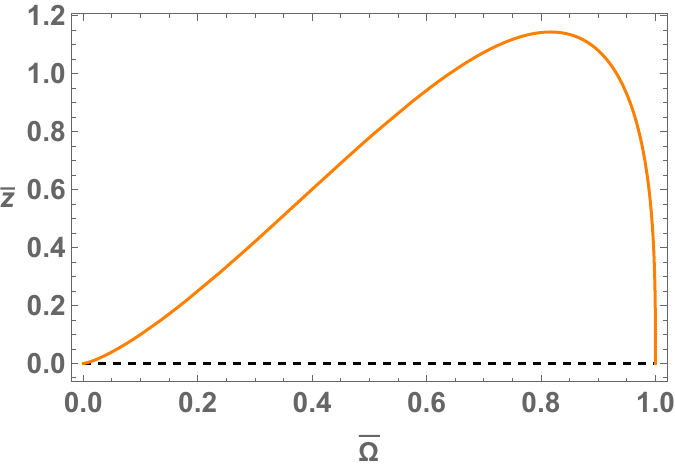}
    \caption{(left) Phase plane for system \eqref{2d-a}-\eqref{2d-b}. (right) Numeric solution of system \eqref{2d-a}-\eqref{2d-b} for the initial conditions $\overline \Omega=0.1$ and $\overline z=0.1$.}
    \label{fig:2}
\end{figure}
After the averaging process \eqref{timeavrg}, the leading terms are
\begin{subequations}
\begin{align}
    & \overline \Omega^{\prime}=\frac{3}{2} \overline \Omega  \left(\overline \Omega^2-1\right), \label{2d-a}\\
    & \overline z^{\prime}=\left(3 \overline \Omega ^2-2\right) \overline z, \label{2d-b}\\
    & \overline v^{\prime}=2 -2 \omega ^2  -3 \overline v \overline \Omega ^2, \quad \overline \Phi^{\prime}=\frac{3 H \left(\omega ^2-1\right)^3 \overline \Omega ^2}{8 \omega ^3}, \quad H^{\prime}=-\frac{3 H \Omega ^2}{2}.
\end{align}
\end{subequations}

Therefore, the leading terms as $H\rightarrow 0$ of the full system \eqref{perts2} are expected to be 
    \begin{align*}
    &\Omega (\tau )= \frac{1}{\sqrt{1+e^{3 \tau +2 c_1}}}, \quad z(\tau )= \frac{c_2 e^{\tau
   }}{1+e^{3 \tau +2 c_1}}, \quad H(\tau )= c_3 e^{-3 \tau /2} \sqrt{1+e^{3
   \tau +2 c_1}}\\
   & y(\tau )=c_4-\frac{2 e^{-2 c_1} \left(\omega
   ^2-1\right) \ln \left(1+e^{3 \tau +2 c_1}\right)}{3 c_3{}^2}, \quad \Phi(\tau)= c_5-\frac{c_3 e^{-3 \tau /2} \left(\omega
   ^2-1\right)^3 \sqrt{1+e^{3 \tau +2 c_1}}}{4 \omega ^3},
\end{align*}
where $c_1,c_2, c_3, c_4$ and $c_5$ are integration constants, we also verify that in the limit $\tau \rightarrow \infty$ both $\overline\Omega (\tau )\rightarrow 0$ and $\overline z(\tau ) \rightarrow 0$ meaning that we are in the presence of long wavelength perturbations. $H (\tau )\rightarrow e^{c_1} c_3, y (\tau ) \simeq c_4-\frac{2 e^{-2 c_1} \left(\omega ^2-1\right) (3 \tau +2
   c_1)}{3 c_3{}^2} \rightarrow \left(\omega ^2-1\right) (-\infty ),  \Phi (\tau )\rightarrow c_5-\frac{e^{c_1} c_3 \left(\omega ^2-1\right)^3}{4 \omega   ^3}$.

System \eqref{2d-a}-\eqref{2d-b} has two equilibrium points 
\begin{enumerate}
    \item $(\overline \Omega, \overline z)=(0,0)$ with eigenvalues $\left\{-2,-\frac{3}{2}\right\}$. This point is a sink.
    \item $(\overline \Omega, \overline z)=(1,0)$ with eigenvalues $\{3,1\}$. This point is a source.
\end{enumerate}
In Fig. \ref{fig:2} on the left is presented a phase plane for system \eqref{2d-a}-\eqref{2d-b}. On the right, we show a numeric solution of system \eqref{2d-a}-\eqref{2d-b} for the initial conditions $\overline \Omega=0.1$ and $\overline z=0.1$.

\section{Bianchi V metric}
\label{Sect:3.4}
We consider the Bianchi type-V metric in the form \cite{coley2003dynamical, Pradhan:2003dp, Pradhan:2004wt, Pradhan:2004ip, Christodoulakis:2005fx, Singh:2007zzg, Singh:2008zzi, Bali:2008zzc, Terzis:2010dk, Sarkar:2014ysa, Ali:2015nha, Mitsopoulos:2019afs, Mahmood:2020swa, Paliathanasis:2023lim}
\begin{align}
ds^2 =-dt^2 +a(t)^2 dx^2 +b(t)^2 e^{2x} (dy^2 +a(t)^4b(t)^{-4}dz^2).    
\end{align}
The Hubble factor is 
\begin{equation}
H=    \frac{\dot{a}}{a}
\end{equation}
The tensor of anisotropies has the form
\begin{align}
    \left(
\begin{array}{ccc}
 0 & 0 & 0 \\
 0 & -\sigma & 0 \\
 0 & 0 & \sigma \\
\end{array}
\right), \quad \sigma=\frac{\dot{a}}{a }-\frac{\dot{b}}{b},
\end{align}
and the 3-Ricci curvature is 
\begin{align}
 & {}^{(3)}R= -6 a^{-2}.
\end{align}

The field equations are: 
\begin{subequations}
\begin{align}
&\ddot\phi+3 H \dot \phi +\phi + \sin\left( \frac{\phi}{f}\right)= 0, \label{KG}\\
&\dot{\rho_m}+3\gamma H\rho_m=0,\\
& \dot{a}= a H, \\
& \dot{H}=-\frac{1}{2} \left(\gamma   \rho _{m}+2 \sigma^2+{\dot{\phi}}^2\right)-\frac{1}{a^2}, 
\\
& \dot{\sigma}= -3 H \sigma,
\end{align}
\end{subequations}
with constraint
\begin{align}\label{FBV}
3 H^2 = \sigma ^2 +\rho_m+\frac{1}{2}{\dot{\phi}}^2 + \frac{\phi^2}{2} +f\left[1- \cos \left(\frac{\phi }{f}\right)\right] + \frac{3}{a^2}.
 \end{align}

 Defining
\begin{equation}
\Omega_k= \frac{1}{a^2 H^2},  \;   \Sigma=\frac{\sigma}{\sqrt{3}H}
\end{equation}
we obtain the full system 
\begin{small}
\begin{subequations}
\begin{align}
\label{fullsyst-a}
  &\dot{H}=-\frac{1}{2} H^2 \left(3 \gamma   \Omega_{m}+6 \Sigma ^2+3 \Omega ^2 \cos (2 (\Phi -t \omega
   ))+3 \Omega ^2+2  \Omega_k\right),\\
\label{fullsyst-b}& \dot{\Omega}=\frac{1}{12} \Bigg(\frac{2 \sqrt{6} \cos
   (\Phi -t \omega )   \sin \left(\frac{\sqrt{6}
   H \Omega  \sin (\Phi -t \omega )}{f \omega
   }\right)}{H}-\frac{6
   \left(\omega ^2-1\right) \Omega  \sin (2 (\Phi -t
   \omega ))}{\omega }  \\
\label{fullsyst-c}& +6 H \Omega  \left(3 \gamma 
   \Omega_m+6 \Sigma ^2+3 \left(\Omega
   ^2-1\right) \cos (2 (\Phi -t \omega ))+3 \Omega
   ^2+2 \Omega_k-3\right)\Bigg),\\
\label{fullsyst-d}& \dot{\Sigma}=\frac{1}{2} H \Sigma  \left(3 \gamma 
   \Omega_m+6 \Sigma ^2+3 \Omega ^2 \cos (2
   (\Phi -t \omega ))+3 \Omega ^2+2 \Omega_k-6\right),\\
\label{fullsyst-e}& \dot{\Omega}_k=H \Omega_k \left(3 \gamma 
   \Omega_m+6 \Sigma ^2+3 \Omega ^2 \cos (2
   (\Phi -t \omega ))+3 \Omega ^2+2  \Omega_k-2\right),\\
\label{fullsyst-f}& \dot{\Phi}=\frac{1}{12} \sin (\Phi -t \omega )
   \Bigg(-\frac{2\sqrt{6}  \sin \left(\frac{\sqrt{6} H \Omega  \sin (\Phi -t \omega
   )}{f \omega }\right)}{H \Omega }+\frac{12 \left(\omega ^2-1\right) \sin (\Phi -t
   \omega )}{\omega }   +36 H \cos (\Phi -t \omega
   )\Bigg),
\end{align}
\end{subequations}
\end{small}
where 
\begin{small}
\begin{align}\label{OmegamBV}
    \Omega_m & =\frac{1}{6} \Bigg(\frac{2 f \cos \left(\frac{\sqrt{6} H \Omega  \sin (\Phi -t
   \omega )}{f \omega }\right)}{H^2} -\frac{2 f}{H^2}-6 \Sigma ^2-\frac{3
   \left(\omega ^2-1\right) \Omega ^2 \cos (2 (\Phi -t \omega ))}{\omega
   ^2}  \nonumber\\  
   & +\left(-\frac{3}{\omega ^2}-3\right) \Omega ^2-6  \Omega_k+6\Bigg).
\end{align}
\end{small}
Taking series around $H=0$ and setting  $\omega ^2>1$ and $f=(\omega^2-1)^{-1}$ we obtain the following system
\begin{subequations}
\label{BV}
\begin{align}
& \dot{H} =-H^2 \left[3 \Sigma ^2+6 \Omega ^2 \cos ^2(\Phi -t \omega )+ \Omega_k+\frac{3 \gamma  (1- \Omega^2 - \Sigma^2 -\Omega_k)}{2}\right],
\\
    & \dot{\mathbf{x}}= H \mathbf{f}(\mathbf{x}, t) + \mathcal{O}(H^2), \;   \mathbf{x}= \left(\Omega, \Sigma, \Omega_k, \Phi \right)^T,
  \\
   & \mathbf{f}(\mathbf{x}, t) \nonumber \\
   & = 
   \begin{bmatrix}
\frac{1}{4}  \left(2 (2-3 \gamma ) \Omega 
   \Omega _k+6 \Omega 
   \left(-\gamma  \left(\Sigma ^2+\Omega ^2-1\right)+2 \Sigma ^2+\left(\Omega ^2-1\right) \cos (2 (\Phi -t \omega
   ))+\Omega ^2-1\right)\right)\\
    \frac{1}{2}  \Sigma  \left(-3
   (\gamma -2) \left(\Sigma ^2-1\right)-3 (\gamma -1) \Omega ^2+(2-3 \gamma ) \Omega _k+3 \Omega ^2 \cos (2 (\Phi
   -t \omega ))\right)\\
   \Omega _k \left(-3 \gamma  \left(\Sigma ^2+\Omega ^2-1\right)+(2-3 \gamma ) \Omega
   _k+6 \Sigma ^2+3 \Omega ^2 \cos (2 (\Phi -t \omega ))+3 \Omega ^2-2\right)\\
   3  \sin (\Phi -t \omega )   \cos (\Phi -t \omega )
   \end{bmatrix}.
\end{align}
\end{subequations}

Replacing $\dot{\mathbf{x}}= H \mathbf{f}(\mathbf{x}, t)$ with $\mathbf{f}(\mathbf{x}, t)$ as defined in the third equation of \eqref{BV} by $\dot{\mathbf{y}}= H  \overline{\mathbf{f}}(\mathbf{y})$ with  $\mathbf{y}= \left(\overline{\Omega},  \overline{\Sigma}, \overline{\Omega}_{k}, \overline{\Phi} \right)^T$ and $\overline{\mathbf{f}}$ as defined by \eqref{timeavrg}, we obtain the averaged system: 
\begin{align} \label{BV_avrg}
\begin{bmatrix}
\dot{H}\\
\dot{\overline{\Omega}}\\
\dot{\overline{\Sigma}}\\
\dot{\overline{\Omega}}_k\\
\dot{\overline{\Phi}}
\end{bmatrix}=
H \begin{bmatrix}
-\frac{1}{2} H  \left(3 \gamma  (1-\overline{\Omega}^2 - \overline{\Sigma}^2 -\overline{\Omega}_k) +6 \overline{\Sigma}^2+3 \overline{\Omega}^2+2 \overline{\Omega}_{k}\right)\\
 \frac{1}{2}   \overline{\Omega } \left(-3 (\gamma -2)  \overline{\Sigma }^2-3 (\gamma -1) \left( \overline{\Omega }^2-1\right)+(2-3
   \gamma )  \overline{\Omega }_k\right) \\
    \frac{1}{2}   \overline{\Sigma } \left(-3 (\gamma -2) \left( \overline{\Sigma }^2-1\right)-3 (\gamma -1)  \overline{\Omega }^2+(2-3
   \gamma )  \overline{\Omega }_k\right) \\
   \overline{\Omega }_k \left(-3 (\gamma -2)  \overline{\Sigma }^2-3 (\gamma -1)  \overline{\Omega }^2-(3 \gamma -2) \left( \overline{\Omega
   }_k-1\right)\right) \\
 0 
\end{bmatrix}
 .
\end{align}
Introducing the new variable  $\tau= \ln a$, it follows the   regular dynamical system: 
\begin{align}\label{guidingBV}
\begin{bmatrix}
{\overline{\Omega}}^{\prime}\\
{\overline{\Sigma}}^{\prime}\\
{\overline{\Omega}}_k^{\prime}
\end{bmatrix}=
\begin{bmatrix}
-\frac{1}{2}   \overline{\Omega}  \left(3 \gamma  \left(\overline{\Sigma} ^2+\overline{\Omega}^2+\overline{\Omega}_{k}-1\right)-6 \overline{\Sigma} ^2-3 \overline{\Omega} ^2-2 \overline{\Omega}_{k}+3\right)\\
\frac{1}{2}  \overline{\Sigma}  \left(-3 \gamma  \left(\overline{\Sigma}^2+\overline{\Omega}
   ^2+\overline{\Omega}_{k}-1\right)+6 \overline{\Sigma} ^2+3 \overline{\Omega} ^2+2 \overline{\Omega}_{k}-6\right)\\
-\overline{\Omega}_k \left(3 \gamma  \left(\overline{\Sigma} ^2+\overline{\Omega}^2+\overline{\Omega}_k-1\right)-6 \overline{\Sigma} ^2-3 \overline{\Omega}^2-2 \overline{\Omega}_{k}+2\right)
\end{bmatrix}
,
\end{align}
defined on the phase space 
\begin{equation}
    \left\{(\overline{\Omega}, \overline{\Sigma}, \overline{\Omega}_k)\in \mathbb{R}^3| \;  \overline{\Omega}^2 + \overline{\Omega}_k + \overline{\Sigma}^2 \leq 1, \overline{\Omega}\geq 0,  0\leq \overline{\Omega}_k\leq 1, -1\leq \overline{\Sigma} \leq 1 \right\}.
\end{equation}
We assume $\lambda\geq 0, 0\leq \gamma \leq 2$.
\subsection{Dynamical system analysis of the averaged system for Bianchi V metric}
\label{analysis}
In Table \ref{TBV}, we summarize the stability analysis for the equilibrium points of the system 
\eqref{guidingBV}. 

  \begin{table}[t]
\begin{center}
\begin{tabular}{|c|c|c|c|c|}
\hline
Label & $({\overline{\Omega}}, \overline{\Sigma}, \overline{\Omega}_{k})$ & Existence &  Eigenvalues  & Stability \\ \hline 
$P_1$ &  $(0, 0, 0)$ & always & $\lbrace \frac{3 (\gamma -2)}{2},\frac{3 (\gamma -1)}{2},3 \gamma -2 \rbrace$ 
&Sink for $0\leq \gamma <2/3$\\&&&& Saddle for $2/3<\gamma<1,\quad 1<\gamma<2$ \\&&&& Non-hyperbolic for $\gamma=2/3,1,2$\\ \hline 
$P_2$ & $(1,0,0)$ & always & $\lbrace -\frac{3}{2},1,3-3 \gamma \rbrace$ & Saddle for $\gamma \neq 1$\\&&&& Non-hyperbolic saddle for $\gamma=1$\\ \hline
$P_3$ & $(0,0,1)$ & always  & $\lbrace -2,-\frac{1}{2},2-3 \gamma  \rbrace$ & Sink for $2/3<\gamma \leq 2$  \\&&&& Saddle for $0\leq \gamma <2/3$ \\&&&& Non-hyperbolic for $\gamma=2/3$\\ \hline
$P_{4,5}$ & $(0,\pm 1, 0)$ & always & $\lbrace 4,\frac{3}{2},6-3 \gamma  \rbrace$ &Source for $0\leq \gamma <2$  \\&&&& Non-hyperbolic for $\gamma=2$  \\ \hline
\end{tabular}
\end{center}\caption{Stability analysis for the equilibrium points of
\eqref{guidingBV}.}
\label{TBV}
\end{table}

\begin{figure}[h!]
    \centering
    \includegraphics[scale=0.6]{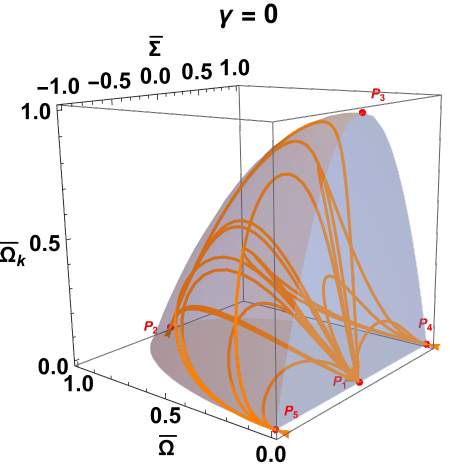}
    \includegraphics[scale=0.6]{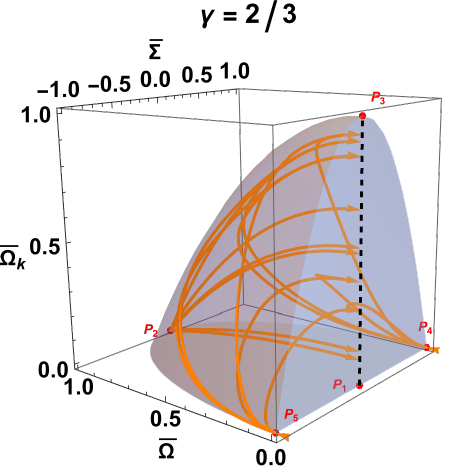}
    \includegraphics[scale=0.6]{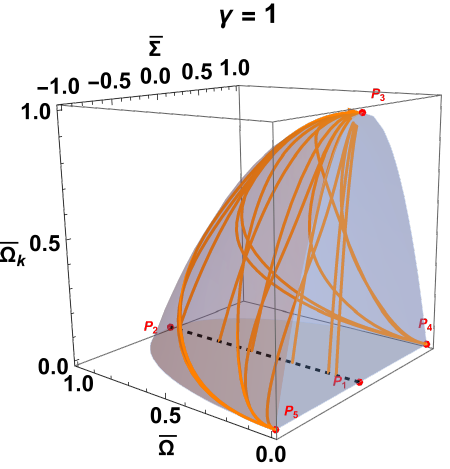}
    \includegraphics[scale=0.6]{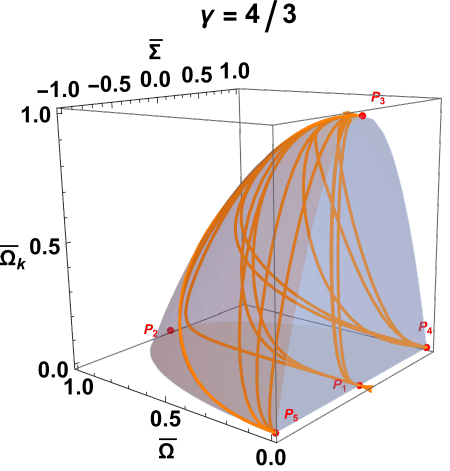}
    \includegraphics[scale=0.6]{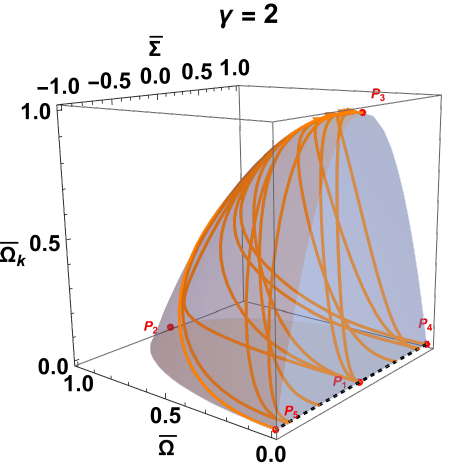}
    \caption{Phase space analysis of system \eqref{guidingBV} for the following values of $\gamma=0,2/3,1,4/3,2$}
    \label{FIG:1}
\end{figure}

They are the following.

\begin{enumerate}
    \item $P_1$ with coordinates  $(\overline{\Omega}, \overline{\Sigma}, \overline{\Omega}_k)=(0,0,0)$ always exists. It represents the flat, matter-dominated FLRW solution. It is 
    \begin{enumerate}
        \item a sink for $0\leq \gamma <2/3$,
        \item a saddle for $2/3<\gamma<1,\quad 1<\gamma<2$,
        \item  Non-hyperbolic for $\gamma=2/3,1,2$.
    \end{enumerate}
    \item $P_2$ with coordinates $(1, 0, 0)$ always exists. It corresponds to the flat scalar field-dominated FLRW solution. It is 
    \begin{enumerate}
        \item a saddle for $\gamma \neq 1$,
        \item Non-hyperbolic saddle for $\gamma=1$.
    \end{enumerate}
    \item $P_3$ with coordinates $(0, 0, 1)$ always  exists. It corresponds to the curvature-dominated (Milne, $k=-1$) solution. It is 
    \begin{enumerate}
        \item a sink for $2/3<\gamma \leq 2$,
        \item a saddle for $0\leq \gamma <2/3$
        \item Non-hyperbolic for $\gamma=2/3$.
    \end{enumerate}
    \item Points $P_{4,5}$ with coordinates $(0, \pm 1, 0)$  always exist. They are anisotropic vacuum solutions. They are 
    \begin{enumerate}
        \item a source for $0\leq \gamma <2$,
        \item Non-hyperbolic for $\gamma=2$. 
    \end{enumerate} 
\end{enumerate}
In figure \ref{FIG:1}, the dynamical behaviour for system \eqref{guidingBV} is depicted for different values of the equation of state parameter $\gamma.$ The behaviour agrees with the stability analysis performed.  Given this, we can formalize the following result
\begin{thm}
The late-time attractors for the Bianchi V metric are
\begin{enumerate}
    \item[a)] The flat matter-dominated FLRW solution $P_1$ for $0\leq \gamma <\frac{2}{3}.$
    \item[b)] The curvature-dominated Milne solution $P_3$ for 
 $\frac{2}{3}<\gamma\leq 2.$ 
\end{enumerate}
\end{thm}

\subsection{Error estimation and numerical integration of solutions for Bianchi V metric}
\label{numerical}

Proceeding in analogous way as in references \cite{Alho:2015cza,Alho:2019pku} we implement a local nonlinear transformation:   
\begin{small}
\begin{align}
&\mathbf{x}_0:=\left(\Omega_{0}, \Sigma_{0}, \Omega_{k0}, \Phi_0\right)^T  \mapsto \mathbf{x}:=\left(\Omega, \Sigma, \Omega_k, \Phi\right)^T \nonumber \\
& \mathbf{x}=\psi(\mathbf{x}_0):=\mathbf{x}_0 + H \mathbf{g}(H, \mathbf{x}_0,t), \label{BVquasilinear211}
\\
& \mathbf{g}(H, \mathbf{x}_0,t)=  \begin{bmatrix}
    g_1(H, \Omega_{0}, \Sigma_{0}, \Omega_{k0}, \Phi_0, t)\\
    g_2(H, \Omega_{0}, \Sigma_{0}, \Omega_{k0}, \Phi_0, t)\\
    g_3(H, \Omega_{0}, \Sigma_{0}, \Omega_{k0}, \Phi_0, t)\\
    g_4(H, \Omega_{0}, \Sigma_{0}, \Omega_{k0}, \Phi_0, t)\\
 \end{bmatrix}.   \label{eqW55}
\end{align}
\end{small}
Taking time derivative in both sides of \eqref{BVquasilinear211} with respect to $t$ we obtain 
\begin{small}
\begin{align}
    & \dot{\mathbf{x}_0}+ \dot{H} \mathbf{g}(H, \mathbf{x}_0,t)  \nonumber \\
    & + H \Bigg(\frac{\partial }{\partial t} \mathbf{g}(H, \mathbf{x}_0,t) + \dot{H} \frac{\partial }{\partial H} \mathbf{g}(H, \mathbf{x}_0,t)    + D_{\mathbf{x}_0} \mathbf{g}(H, \mathbf{x}_0,t) \cdot \dot{\mathbf{x}_0}\Bigg)   = \dot{\mathbf{x}}, \label{EQT56}
    \end{align}
    \end{small}
    where 
    \begin{equation}
        D_{\mathbf{x}_0} \mathbf{g}(H, \mathbf{x}_0,t)=  \begin{bmatrix}
             \frac{\partial g_1}{\partial \Omega_0}  &  \frac{\partial g_1}{\partial \Sigma_0} & \frac{\partial g_1}{\partial \Omega_k} &  \frac{\partial g_1}{\partial \Phi_0}\\
             \frac{\partial g_2}{\partial \Omega_0}  &  \frac{\partial g_2}{\partial \Sigma_0} & \frac{\partial g_2}{\partial \Omega_k} &  \frac{\partial g_2}{\partial \Phi_0}\\
                \frac{\partial g_3}{\partial \Omega_0} &  \frac{\partial g_3}{\partial \Sigma_0} & \frac{\partial g_3}{\partial \Omega_k} &  \frac{\partial g_3}{\partial \Phi_0}\\
                   \frac{\partial g_4}{\partial \Omega_0}  &  \frac{\partial g_4}{\partial \Sigma_0} & \frac{\partial g_4}{\partial \Omega_k} &  \frac{\partial g_4}{\partial \Phi_0}\\
        \end{bmatrix}
    \end{equation}    
is the  Jacobian matrix of $\mathbf{g}(H, \mathbf{x}_0,t)$ for the vector  $\mathbf{x}_0$.  The function $\mathbf{g}(H, \mathbf{x}_0,t)$ is conveniently chosen. 
\newline By substituting \eqref{equx} and \eqref{BVquasilinear211} in \eqref{EQT56} we obtain 
\begin{small}
\begin{align}
       & \Bigg(\mathbf{I}_4 + H D_{\mathbf{x}_0} \mathbf{g}(H, \mathbf{x}_0,t)\Bigg) \cdot \dot{\mathbf{x}_0}= H \mathbf{f}(\mathbf{x}_0 + H \mathbf{g}(H, \mathbf{x}_0,t),t) \nonumber \\
       & -H \frac{\partial }{\partial t} \mathbf{g}(H, \mathbf{x}_0,t) -\dot{H} \mathbf{g}(H, \mathbf{x}_0,t) -H \dot{H} \frac{\partial }{\partial H} \mathbf{g}(H, \mathbf{x}_0,t), 
\end{align}
\end{small}
where 
$\mathbf{I}_4$ is the $4\times 4$ identity matrix.
                
Then we obtain 
\begin{widetext}
  \begin{small}  
  \begin{align}
 & \dot{\mathbf{x}_0} = \Bigg(\mathbf{I}_4 + H D_{\mathbf{x}_0} \mathbf{g}(H, \mathbf{x}_0,t)\Bigg)^{-1} \cdot \Bigg(H \mathbf{f}(\mathbf{x}_0 + H \mathbf{g}(H, \mathbf{x}_0,t),t)-H \frac{\partial }{\partial t} \mathbf{g}(H, \mathbf{x}_0,t)  \nonumber \\
 & \quad\quad\quad \quad\quad\quad\quad\quad\quad\quad\quad\quad \quad\quad\quad\quad\quad\quad    -\dot{H} \mathbf{g}(H, \mathbf{x}_0,t) -H \dot{H} \frac{\partial }{\partial H} \mathbf{g}(H, \mathbf{x}_0,t)\Bigg).  
\end{align}
\end{small}
Using eq. \eqref{EQ:81b}, we have $ \dot{H}= \mathcal{O}(H^2)$. Hence,
\begin{small}
\begin{align}
    & \dot{\mathbf{x}_0} = \underbrace{\Bigg(\mathbf{I}_4 - H D_{\mathbf{x}_0} \mathbf{g}(0, \mathbf{x}_0,t) +  \mathcal{O}(H^2)\Bigg)}_{4\times 4 \: \text{matrix}} \cdot \underbrace{\Bigg(H \mathbf{f}(\mathbf{x}_0, t)-H \frac{\partial }{\partial t} \mathbf{g}(0, \mathbf{x}_0,t) +   \mathcal{O}(H^2)\Bigg)}_{4\times 1 \; \text{vector}} \nonumber \\
    & = \underbrace{H \mathbf{f}(\mathbf{x}_0, t)-H \frac{\partial }{\partial t} \mathbf{g}(0, \mathbf{x}_0,t) +   \mathcal{O}(H^2)}_{4\times 1 \; \text{vector}}.\label{eqT59}
    \end{align} 
    \end{small}
\end{widetext}
\noindent The strategy is to use eq. \eqref{eqT59} for choosing conveniently $\frac{\partial }{\partial t} \mathbf{g}(0, \mathbf{x}_0,t)$ to prove that 
\begin{align}
 & \dot{\Delta\mathbf{x}_0}= -H G(\mathbf{x}_0, \overline{\mathbf{x}}) +   \mathcal{O}(H^2), \label{EqY60}
  \end{align}
where $\overline{\mathbf{x}}=(\overline{\Omega}, \overline{\Sigma},  \overline{\Omega}_k, \overline{\Phi})^T$ and  $\Delta\mathbf{x}_0=\mathbf{x}_0 - \overline{\mathbf{x}}$. The function $G(\mathbf{x}_0, \overline{\mathbf{x}})$ is unknown at this stage. 
\newline 
By construction, we neglect dependence of $\partial g_i/ \partial t$ and $g_i$ on $H$, i.e., assume $\mathbf{g}=\mathbf{g}(\mathbf{x}_0,t)$ because the dependence of $H$ is dropped out along with higher order terms eq. \eqref{eqT59}. Next, we solve a partial differential equation  for $\mathbf{g}(\mathbf{x}_0,t)$ given by:  
\begin{align}
     & \frac{\partial }{\partial t} \mathbf{g}(\mathbf{x}_0,t) = \mathbf{f}(\mathbf{x}_0, t) - \overline{\mathbf{f}}(\overline{\mathbf{x}}) + G(\mathbf{x}_0, \overline{\mathbf{x}}). \label{eqT60}
\end{align}
\noindent  where we have considered $\mathbf{x}_0$, and $t$ as independent variables. 
\newline
The right-hand side of \eqref{eqT60} is almost periodic of period $L=\frac{2\pi}{\omega}$ for large times. Then, implementing the average process \eqref{timeavrg} on right hand side of \eqref{eqT60}, where slow-varying dependence of quantities $\Omega_{0}, \Sigma_{0}, \Omega_{k0},  \Phi_0$ and  $\overline{\Omega}, \overline{\Omega}_k,  \overline{\Sigma}, \overline{\Phi}$  on $t$ are ignored through  averaging process, we obtain \begin{align}
    & \frac{1}{L}\int_0^{L} \Bigg[\mathbf{f}(\mathbf{x}_0, s) - \overline{\mathbf{f}}(\overline{\mathbf{x}}) +G(\mathbf{x}_0, \overline{\mathbf{x}}) \Bigg] ds   = \overline{\mathbf{f}}( {\mathbf{x}}_0)-\overline{\mathbf{f}}(\overline{\mathbf{x}} )+G(\mathbf{x}_0, \overline{\mathbf{x}}). \label{newaverage}
\end{align}
Defining 
\begin{equation}
  G(\mathbf{x}_0, \overline{\mathbf{x}}):=  -\left(\overline{\mathbf{f}}( {\mathbf{x}}_0)-\overline{\mathbf{f}}(\overline{\mathbf{x}})\right)
\end{equation} the average \eqref{newaverage} is zero so that $\mathbf{g}(\mathbf{x}_0,t)$ is bounded.
\newline 
Finally, eq. \eqref{EqY60} transforms to 
\begin{align}
 & \dot{\Delta\mathbf{x}_0}= H \left(\overline{\mathbf{f}}( {\mathbf{x}}_0)-\overline{\mathbf{f}}(\overline{\mathbf{x}})\right) +   \mathcal{O}(H^2)  \label{EqY602}
  \end{align}
and eq. \eqref{eqT60} 
is simplified to 
\begin{align}
     & \frac{\partial }{\partial t} \mathbf{g}(\mathbf{x}_0,t) = \mathbf{f}(\mathbf{x}_0, t) - \overline{\mathbf{f}}( \mathbf{x}_0). \label{eqT602}
\end{align}

\begin{lem}[\textbf{Gronwall's Lemma (Integral form)}]
\label{Gronwall}
Let be $\xi(t)$ a nonnegative function, summable over  $[0,T]$ which satisfies almost everywhere the integral inequality $$\xi(t)\leq C_1 \int_0^t \xi(s)ds +C_2, \;  C_1, C_2\geq 0.$$
       Then, 
     $\xi(t)\leq C_2  e^{C_1 t},$
        almost everywhere for $t$ in $0\leq t\leq T$.
 In particular, if    
     $$\xi(t)\leq C_1 \int_0^t \xi(s)ds, \;  C_1\geq 0$$
        almost everywhere for $t$ in $0\leq t\leq T$. Then,  $
           \xi \equiv 0$  
        almost everywhere for $t$ in $0\leq t\leq T$.
 \end{lem}
 
\begin{lem}[Mean value theorem]
\label{lemma6}
 Let $U \subset \mathbb{R}^n$ be open, $\mathbf{f}: U \rightarrow \mathbb{R}^m$ continuously differentiable, and $\mathbf{x}\in U$, $\mathbf{h}\in \mathbb{R}^m$ vectors such that the line segment $\mathbf{x}+z \; \mathbf{h}$,  $0 \leq z \leq 1$ remains in $U$. Then we have:
\begin{equation}
    \mathbf{f}(\mathbf{x}+\mathbf{h})-\mathbf{f}(\mathbf{x}) = \left (\int_0^1 D\mathbf{f}(\mathbf{x}+z \; \mathbf{h})\,dz\right)\cdot \mathbf{h},
\end{equation} where  $D \mathbf{f}$ denotes the Jacobian matrix of $\mathbf{f}$ and the integral of a matrix is to be understood componentwise.
\end{lem}
 Theorem \ref{BVLFZ11} establish the existence of the vector \eqref{eqW55}. 
\begin{thm}
\label{BVLFZ11} Let $H, \overline{\Omega}, \overline{\Sigma}, \overline{\Omega}_k, \overline{\Phi}$  be  defined  functions that  satisfy  averaged  equations \eqref{BV_avrg}. Then, there exist continuously differentiable functions $g_1, g_2, g_3$ and $g_4$,  such that   $\Omega, \Sigma, \Omega_k$ and $\Phi$  are locally given by \eqref{BVquasilinear211}, where $\Omega_{0}, \Sigma_{0}, \Omega_{k0}, \Phi_0$ are order zero approximations of them as $H\rightarrow 0$. Then,  functions $\Omega_{0}, \Sigma_{0}, \Omega_{k0}, \Phi_0$ and averaged solutions $\overline{\Omega}, \overline{\Omega}_k,  \overline{\Sigma}, \overline{\Phi}$  have the same limit as $t\rightarrow \infty$. 
Setting $\Sigma=\Sigma_0=0$, analogous results for the negatively curved FLRW model are derived. 
\end{thm}

\textbf{Proof of Theorem \ref{BVLFZ11} }.
Defining $Z= 1-\Omega ^2-\Sigma ^2- \Omega_{k}$ it follows 
from 
\begin{align*}
   & \dot{Z}= -Z H \left(3 (\gamma -2) \Sigma ^2+3 (\gamma -1) \Omega ^2+(3 \gamma -2) \Omega _k-3 \Omega ^2 \cos (2 (\Phi -t
   \omega ))\right)+\mathcal{O}\left(H^2\right)\end{align*}  that the sign of $1-\Omega ^2-\Sigma ^2- \Omega_{k}$
 is invariant as $H\rightarrow 0$. 
 From  equation  \eqref{EQ:81b}
it follows that $H$ is a  monotonic decreasing function of  $t$  if $0<\Omega^2+\Sigma^2+{\Omega_k}<1$. 
 These allow us to define recursively  bootstrapping sequences 
\begin{align}
    & \left\{\begin{array}{c}
       t_0=t_{*}   \\ \\
        H_0=H(t_{*}) 
    \end{array}\right., \quad    
    \left\{\begin{array}{c}
       {t_{n+1}}= {t_{n}} +\frac{1}{H_n}   \\ \\
       H_{n+1}= H(t_{n+1})  
    \end{array}\right.,
\end{align}
such that  $\lim_{n\rightarrow \infty}H_n=0$ y $\lim_{n\rightarrow \infty} t_n=\infty$.
\noindent 
Given  expansions \eqref{AppBIIIquasilinear211}, eqs. \eqref{eqT59} become
\begin{small}
\begin{subequations}
\begin{align}
 & \dot{\Omega}_0=  -\left(\frac{\partial g_1}{\partial t} + \frac{3}{2} \Omega_{0} \left(1-\Omega_{0}^2\right) \cos (2 (\Phi_0-t \omega)) \right) H +\mathcal{O}\left(H^2\right), \\
 & \dot{\Sigma}_0=   -\left(\frac{\partial g_2}{\partial t} -\frac{3}{2} \Sigma_{0} \Omega_{0}^2 \cos (2 (\Phi_0-t \omega)) \right) H +\mathcal{O}\left(H^2\right), \\
 & \dot{\Omega}_{k0}= -\left(\frac{\partial g_3}{\partial t} -  3 \Omega_{0}^2 \Omega_{k0} \cos (2 (\Phi_0-t \omega ))\right)H +\mathcal{O}\left(H^2\right),
\\
 & \dot{\Phi}_0=-\left(\frac{\partial g_4}{\partial t} + \frac{3}{2} \sin (2 (t  \omega - \Phi_0))\right)
   H+\mathcal{O}\left(H^2\right).
\end{align}
\end{subequations}
\end{small}
Furthermore, eqs. \eqref{eqT602} become
\begin{small}
\begin{subequations}
\label{eq28}
\begin{align}
& \frac{\partial g_1}{\partial t}=- \frac{3}{2} \Omega_{0} \left(1-\Omega_{0}^2\right) \cos (2 (\Phi_0-t \omega)),
\\
& \frac{\partial g_2}{\partial t}=\frac{3}{2} \Sigma_{0} \Omega_{0}^2 \cos (2 (\Phi_0-t \omega)),
\\
& \frac{\partial g_3}{\partial t}=3 \Omega_{0}^2 \Omega_{k0} \cos (2 (\Phi_0-t \omega )),
\\
&  \frac{\partial g_4}{\partial t}=\frac{3}{2} \sin (2 (\Phi_0-t \omega)).
\end{align}
\end{subequations}
\end{small}
Then, explicit expressions of the $g_i, i=1, \ldots, 4$ are found by  integration of \eqref{eq28}:
\begin{small}
\begin{subequations}
\begin{align}
& g_1(H , \Omega_{0}, \Sigma_{0}, \Omega_{k0}, \Phi_0, t) =   -\frac{3 \Omega_0 \left(\Omega_0^2-1\right) \sin (2 (\Phi_0-t \omega ))}{4 \omega}  +C_1(\Omega_{0}, \Sigma_{0}, \Omega_{k0}, \Phi_0), 
\\
& g_2(H , \Omega_{0}, \Sigma_{0}, \Omega_{k0}, \Phi_0, t)   = -\frac{3 \Sigma_0 \Omega_0^2 \sin
   (2 (\Phi_0-t \omega ))}{4 \omega } +C_2(\Omega_{0}, \Sigma_{0}, \Omega_{k0}, \Phi_0),
\\
& g_3(H , \Omega_{0}, \Sigma_{0}, \Omega_{k0}, \Phi_0, t)  =  -\frac{3 \Omega_0^2 \Omega_{k0} \sin (2 (\Phi_0-t \omega ))}{2 \omega } +C_3(\Omega_{0}, \Sigma_{0}, \Omega_{k0}, \Phi_0),
\\
&g_4(H , \Omega_{0}, \Sigma_{0}, \Omega_{k0}, \Phi_0, t)   =  \frac{3 \cos (2 (\Phi_0-t \omega ))}{4 \omega } +C_4(\Omega_{0}, \Sigma_{0}, \Omega_{k0}, \Phi_0), 
\end{align}
\end{subequations}
\end{small}
where we can set four integration functions $C_i(\Omega_{0}, \Sigma_{0}, \Omega_{k0}, \Phi_0), i = 1,2,3,4$ to zero. Functions   $g_i, i=1, \ldots, 4$ are continuously differentiable, such that their partial derivatives are bounded on $t\in [t_n, t_{n+1}]$.  
Let    $\Delta \Omega_0= \Omega_0 - \overline{\Omega}, \;  \Delta \Sigma_0= \Sigma_0 - \overline{\Sigma},  \;  \Delta \Omega_{k0}= \Omega_{k0}- {\overline{\Omega}_k}, \;  \Delta \Phi_0= \Phi_0 - \overline{\Phi}$ be defined such that 
$\Omega_0(t_n)=\overline{\Omega}(t_n)= {\Omega_{n}},   \;   \Sigma_0(t_n)=\overline{\Sigma}(t_n)= {\Sigma}_{n}, \;  \Omega_{k0}(t_n)={\overline{\Omega}_k}(t_n)= {\Omega_k}_{n},\;  \Phi_0(t_n)=\overline{\Phi}(t_n)= {\Phi}_{n},$
with 
$0<\Omega(t_n) +\Sigma(t_n)^2 +{\Omega_k}(t_n)^2 <1.$

Keeping the terms of second order in $H$,  system \eqref{EqY602} becomes
\begin{small}
\begin{subequations}
\label{eqA8}
\begin{align}
   & \dot{\Delta \Omega_0}= H \Bigg(\frac{1}{2} \Omega_{0} \left(-3 \gamma  \left(\Sigma_0^2+\Omega_{0}^2+\Omega_{k0}-1\right)+6 \Sigma_0^2+3 \Omega_{0}^2+2 \Omega_{k0}-3\right)
   \nonumber \\
   & -\frac{1}{2} \overline{\Omega } \left(-3 \gamma  \left(\overline{\Sigma }^2+\overline{\Omega
   }^2+\overline{\Omega}_{k}-1\right)+6 \overline{\Sigma }^2+3 \overline{\Omega }^2+2 \overline{\Omega}_{k}-3\right)\Bigg) \nonumber \\
   & +\frac{\Omega_0 \sin (2 (\Phi_0-t \omega ))}{8 \omega ^3}  H^2 \Bigg[\omega ^2\Big(-18 \left(\Omega_0^2-1\right) \Sigma_0^2+\Omega_0^2 \left(-2 \omega ^4+6 \omega ^2-9 \Omega_0^2-6 \Omega_{k0}+3\right) \nonumber \\
   & +6 \Omega_{k0}+9
   \gamma  \left(\Omega_0^2-1\right) \left(\Sigma_0^2+\Omega_0^2+\Omega_{k0}-1\right)\Big)  \nonumber \\
   & +2 \Omega_0^2+\left(\omega ^2 \left(-36 \Omega_0^4+\left(2 \omega
   ^4-6 \omega ^2+33\right) \Omega_0^2+9\right)-2 \Omega_0^2\right) \cos (2 (\Phi_0-t \omega
   ))\Bigg] +\mathcal{O}(H^3), 
\end{align}
\begin{align}
   & \dot{\Delta \Sigma_0}=H \Bigg(\frac{1}{32} \overline{\Sigma } \left(48 (\gamma -2) \left(\overline{\Sigma }^2-1\right)+48
   (\gamma -1) \overline{\Omega }^2-16 (2-3 \gamma ) \overline{\Omega}_{k}\right)  \nonumber \\
   & +\frac{1}{32} \Sigma_0
   \left(-48 (\gamma -2) \left(\Sigma_0^2-1\right)-48 (\gamma -1) \Omega_{0}^2+16 (2-3 \gamma )
   \Omega_{k0}\right)\Bigg) \nonumber \\
   & - \frac{3 \Sigma_0 \Omega_0^2 \sin (2
   (\Phi_0-t \omega ))}{8 \omega } H^2 \Bigg[-3 \gamma  \left(\Sigma_0^2+\Omega_0^2+\Omega_{k0}-1\right) \nonumber \\
   & +6 \Sigma_0^2+12 \Omega_0^2 \cos (2 (\Phi_0-t \omega ))+3 \Omega_{0}^2+2 \Omega_{k0}\Bigg] +\mathcal{O}(H^3),
\\
   & \dot{\Delta \Omega_{k0}}=H \Bigg(\Omega_{k0} \left(-3 \gamma  \left(\Sigma_{0}^2+\Omega_{0}^2+\Omega_{k0}-1\right)+6 \Sigma_0^2+3 \Omega_{0}^2+2 \Omega_{k0}-2\right) \nonumber \\
   & -\overline{\Omega}_{k} \left(-3 \gamma  \left(\overline{\Sigma }^2+\overline{\Omega
   }^2+\overline{\Omega}_{k}-1\right)+6 \overline{\Sigma }^2+3 \overline{\Omega }^2+2 \overline{\Omega}_k-2\right)\Bigg)\nonumber \\
   & -\frac{3 \Omega_0^2 \Omega_{k0} \sin (2 (\Phi_0-t \omega )) }{4 \omega } H^2 \Big[-3 \gamma  \left(\Sigma_0^2+\Omega_0^2+\Omega_{k0}-1\right) \nonumber \\
   & +6 \Sigma_0^2+15 \Omega_0^2 \cos (2 (\Phi_0-t \omega ))+3 \Omega_0^2+2
   \Omega_{k0}\Big] +\mathcal{O}(H^3) ,
\\
   & \dot{\Delta \Phi_0}=\frac{1}{16 \omega ^3} H^2 \Bigg[ -6 \left(\omega ^2-1\right)^3 \overline{\Omega }^2 \nonumber \\
   & -2 \cos (2 (\Phi_0-t \omega )) \left(\omega ^2 \left(9 \gamma  \left(\Sigma_0^2+\Omega_0^2+\Omega_{k0}-1\right)-18 \Sigma_0^2+\left(4 \left(\omega ^2-3\right) \omega ^2+3\right) \Omega_0^2-6
   \Omega_{k0}\right)-4 \Omega_0^2\right) \nonumber \\
   & +\left(18 \omega ^2+\left(2 \omega ^6-6 \omega ^4+15 \omega
   ^2-2\right) \Omega_0^2\right) \cos (4 (\Phi_0-t \omega )) \nonumber \\
   & +18 \omega ^2+3 \left(2 \omega ^6-6
   \omega ^4+9 \omega ^2-2\right) \Omega_0^2\Bigg] +\mathcal{O}(H^3) \label{A11}
\end{align}
\end{subequations}
\end{small}
\noindent 
Denoting $\mathbf{x}_0=(\Omega_0, \Sigma_0, \Omega_{k0})^T$, $\overline{\mathbf{x}}=(\overline{\Omega}, \overline{\Sigma}, \overline{\Omega}_{k})^T$ the system \eqref{eqA8} can be written as  
\begin{align*}
 & \dot{\Delta\mathbf{x}_0}= H \left(\overline{\mathbf{f}}( {\mathbf{x}}_0)-\overline{\mathbf{f}}(\overline{\mathbf{x}})\right) + H^2 \mathbf{f}^{[2]}( {\mathbf{x}}_0, \overline{\mathbf{x}}) +\mathcal{O}(H^3), \\
  \end{align*}
plus eq. \eqref{A11}, which can be written symbolically as
\begin{align}
     & \dot{\Delta \Phi_0}=  H^2 \mathbf{\Phi}^{[2]}( {\mathbf{x}}_0,\Phi_0, \overline{\mathbf{x}}, \overline{\Phi}) +\mathcal{O}(H^3),
\end{align} where the vector function $\overline{\mathbf{f}}$ 
is given explicitly (last row corresponding to $\Delta{\Phi}_0$ was omitted) by: 
\begin{small}
\begin{align*}
    \overline{\mathbf{f}}(x, y, z)=\left(
\begin{array}{c}
 \frac{1}{2} x \left(-3 (\gamma -1) \left(x^2-1\right)-3 (\gamma -2) y^2+(2-3 \gamma ) z\right) \\
 \frac{1}{2} y \left(-3 (\gamma -1) x^2-3 (\gamma -2) \left(y^2-1\right)+(2-3 \gamma ) z\right) \\
 z \left(-3 (\gamma -1) x^2-3 (\gamma -2) y^2-(3 \gamma -2) (z-1)\right) \\
\end{array}
\right).
\end{align*}
\end{small}
It is a vector function with polynomial components in variables $(x, y, z)$. Therefore, it is continuously differentiable in all its components.  

Let be $\Delta\mathbf{x}_0(t)= (\Omega_0-\overline{\Omega},  {\Sigma_0}-  \overline{\Sigma}, \overline{\Omega}_k - \Omega_{k0})^T$ with $0\leq |\Delta\mathbf{x}_0|:=\max \left\{|\Omega_0-\overline{\Omega}|,  
|{\Sigma_0}-  \overline{\Sigma}|, |\Omega_{k0} - \overline{\Omega}_k|\right\}$  finite in the closed interval $[t_n,t]$. Using same initial conditions for $\mathbf{x}_0$ and $\overline{\mathbf{x}}$ we obtain by integration: 
\begin{align*}
 \Delta\mathbf{x}_0(t) = \int_{t_n}^t \dot{\Delta\mathbf{x}_0} d s =  \int_{t_n}^t \left(H \left(\overline{\mathbf{f}}( {\mathbf{x}}_0)-\overline{\mathbf{f}}(\overline{\mathbf{x}})\right) +   H^2 \mathbf{f}^{[2]}( {\mathbf{x}}_0, \overline{\mathbf{x}}) +\mathcal{O}(H^3)\right) ds. 
\end{align*}

Using Lemma \ref{lemma6} we have 
\begin{equation}
   \overline{\mathbf{f}}( {\mathbf{x}}_0(s))-\overline{\mathbf{f}}(\overline{\mathbf{x}}(s)) = \underbrace{\left (\int_0^1 D   \overline{\mathbf{f}}\left(\overline{\mathbf{x}}(s)+ z \; \left({\mathbf{x}}_0(s) - \overline{\mathbf{x}}(s)\right)\right)\,d z\right)}_{\mathbf{A}(s)}\cdot \left({\mathbf{x}}_0(s) - \overline{\mathbf{x}}(s)\right),
\end{equation} where  $D\overline{\mathbf{f}}$ denotes the Jacobian matrix of $\overline{\mathbf{f}}$ and the integral of a matrix is to be understood componentwise, where the matrix
$\mathbf{A}=\left(a_{i j}\right)$ has polynomial components $a_{i j} \left(\Omega_{0}, \Sigma_{0}, \Omega_{k0}, \Phi_0, \overline{\Omega},  \overline{\Omega}_k,  \overline{\Sigma}, \overline{\Phi}\right)$. 
Taking   sup norm
$ |\Delta\mathbf{x}_0|=\max \left\{|\Omega_0-\overline{\Omega}|, |{\Sigma_0}-  \overline{\Sigma}|, |\Omega_{k0}-\overline{\Omega}_k| \right\}$ and the sup norm of a matrix
${|} \mathbf{A} {|}$ defined by $\max\{|a_{ij}|:  i=1,2,3, j=1,2,3\}$, 
we have 
\begin{equation*}
    \Big{|} \mathbf{A}(s) \cdot \Delta\mathbf{x}_0(s) \Big{|}\leq 3 \Big{|} \mathbf{A}(s) \Big{|} \Big{|}\Delta\mathbf{x}_0(s)\Big{|}, \quad \forall s\in [t_n, t_{n+1}].
\end{equation*}
By continuity of polynomials $a_{i j} \left(\Omega_{0}, \Sigma_{0}, \Omega_{k0}, \Phi_0, \overline{\Omega},  \overline{\Omega}_k,  \overline{\Sigma}, \overline{\Phi}\right)$ and by continuity of  functions $\Omega_{0}, \Sigma_{0}, \Omega_{k0}, \Phi_0$ and  $\overline{\Omega}, \overline{\Omega}_k,  \overline{\Sigma}, \overline{\Phi}$  in $[t_n, t_{n+1}]$ the following finite constants are found:
\begin{equation*}
    L_1= 3 \max_{t\in[t_n,t_{n+1}]} \Big{|} \mathbf{A} (t)\Big{|}, 
 M_1= \max_{t\in[t_{n},t_{n+1}]} \Bigg{|} \mathbf{f}^{[2]}({\mathbf{x}}_0(t), \overline{\mathbf{x}}(t)) \Bigg{|},
M_2= \max_{t\in[t_{n},t_{n+1}]}\Bigg{|} \mathbf{\Phi}^{[2]}({\mathbf{x}}_0,\Phi_0, \overline{\mathbf{x}}, \overline{\Phi})\Bigg{|}, 
\end{equation*}
such that for all $t\in[t_n, t_{n+1}]$: 
\begin{align*}
 & \Big{|}\Delta\mathbf{x}_0(t) \Big{|} = \Bigg{|} \int_{t_n}^t \dot{\Delta\mathbf{x}_0} d s \Bigg{|} = \Bigg{|} \int_{t_n}^t \Bigg(H \left(\overline{\mathbf{f}}( {\mathbf{x}}_0)-\overline{\mathbf{f}}(\overline{\mathbf{x}})\right) +   H^2 \mathbf{f}^{[2]}( {\mathbf{x}}_0, \overline{\mathbf{x}}) + \mathcal{O}(H^3)\Bigg) ds  \Bigg{|}\nonumber \\
 & \leq H_n \int_{t_n}^t \Big{|}  \overline{\mathbf{f}}( {\mathbf{x}}_0)-\overline{\mathbf{f}}(\overline{\mathbf{x}}) \Big{|} ds +   M_1 H_n^2 (t-t_n)
  \leq H_n \int_{t_n}^t  \Big{|} \mathbf{A}(s) \cdot \Delta\mathbf{x}_0(s) \Big{|} ds  +   M_1 H_n^2 (t-t_n)\nonumber \\
& \leq L_1 H_n \int_{t_n}^t  \Big{|}
 \Delta\mathbf{x}_0(s) \Big{|} ds +   M_1 H_n^2 (t-t_n) \leq  L_1 H_n \int_{t_n}^t  \Big{|}  \Delta\mathbf{x}_0(s) \Big{|} ds +   M_1 H_n,
\end{align*}
due to  $t-t_n\leq {t_{n+1}}- {t_{n}} =\frac{1}{H_n}$.
Using   Gronwall's Lemma \ref{Gronwall}, we have for $t \in[t_n, t_{n+1}]$: 
\begin{align*}
 & \Big{|} \Delta \mathbf{x}_0(t)  \Big{|} \leq   M_1  H_n   e^{L_1  H_n(t-t_n)} \leq    M_1  {H_n}e^{L_1}, \; \text{due to}\; t-t_n\leq {t_{n+1}}- {t_{n}} =\frac{1}{H_n}.
 \end{align*}
 Then, 
 \begin{align*}
& \Big{|} \Delta \Omega_0(t) \Big{|} \leq    M_1 e^{L_1} {H_n}, \;  \Big{|} \Delta \Sigma_0(t) \Big{|} \leq     M_1 e^{L_1} {H_n}, \; \Big{|} \Delta \Omega_{k0}(t) \Big{|} \leq   M_1 e^{L_1} {H_n}.
\end{align*}
Furthermore, from eq. \eqref{A11} we have
\begin{align*}
& |\Delta \Phi_0(t)|= |\Phi_0(t)- \overline{\Phi}(t)| = \Bigg{|} \int_{t_n}^{t} \left( \dot{\Phi_0}(s)- \dot{\overline{\Phi}}(s)\right) d s\Bigg{|}  = \Bigg{|} \int_{t_n}^{t} \left[ H^2 \mathbf{\Phi}^{[2]}( {\mathbf{x}}_0,\Phi_0, \overline{\mathbf{x}}, \overline{\Phi}) +\mathcal{O}(H^3) \right] d s\Bigg{|} \nonumber \\
  &  \leq  M_2 H_n^2 (t-t_n) + \Big{|}\mathcal{O}({ {H_n}}^3)\Big{|}   \leq   M_2 H_n, \; \text{due to}\; t-t_n\leq {t_{n+1}}- {t_{n}} =\frac{1}{H_n}.
   \end{align*}
Finally, taking the limit as $n\rightarrow \infty$, we obtain $H_n\rightarrow  0$. Then,
as $H_n \rightarrow 0$,   functions $\Omega_{0}, \Sigma_{0}, \Omega_{k0}, \Phi_0$ and  $\overline{\Omega},  \overline{\Sigma}, \overline{\Omega}_k, \overline{\Phi}$  have the same limit as $\tau\rightarrow \infty$.
\newline 
Setting $\Sigma=\Sigma_0=0$, analogous results for the negatively curved FLRW model are derived. 
$\square$

\begin{figure}[h!]
    \centering
    \includegraphics[scale=0.6]{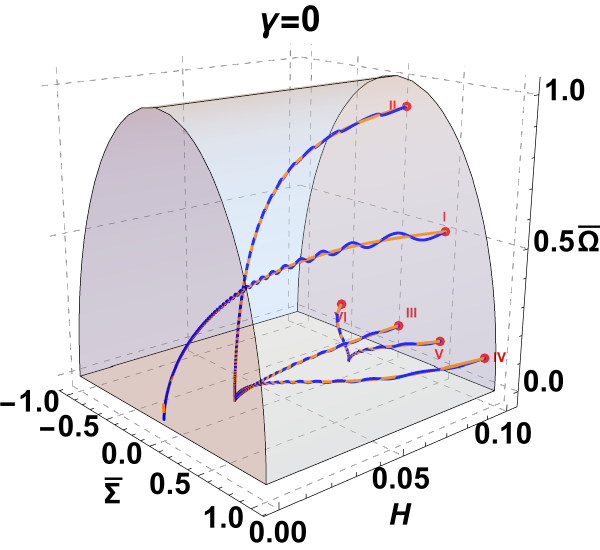}
    \includegraphics[scale=0.6]{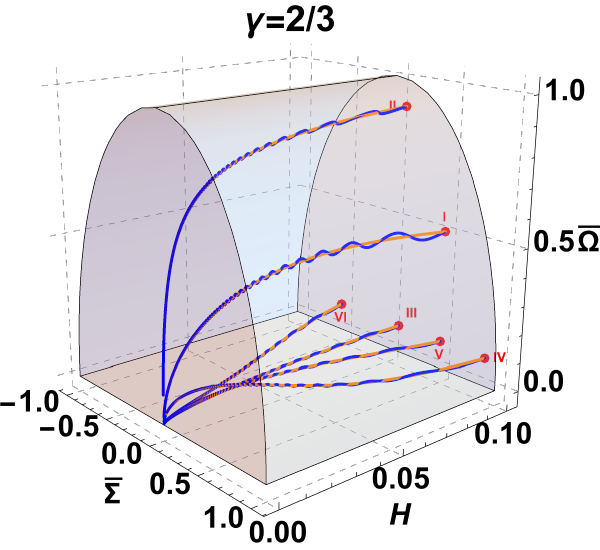}
    
    \includegraphics[scale=0.6]{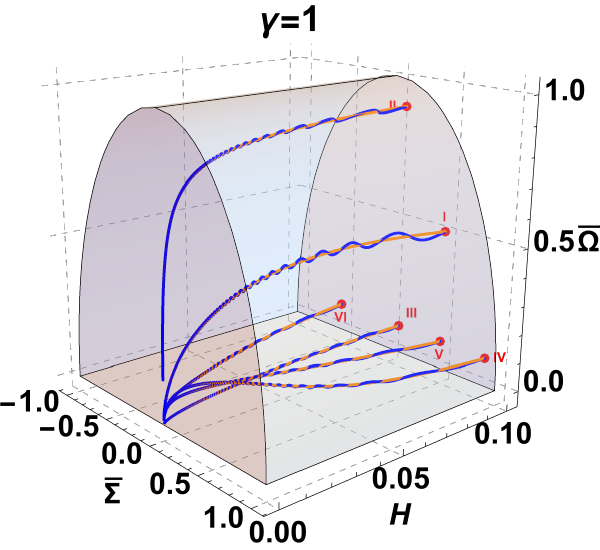}
    \includegraphics[scale=0.6]{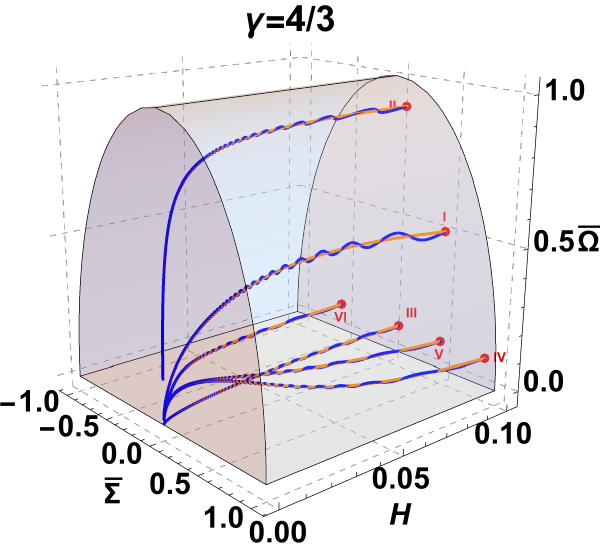}
    \includegraphics[scale=0.6]{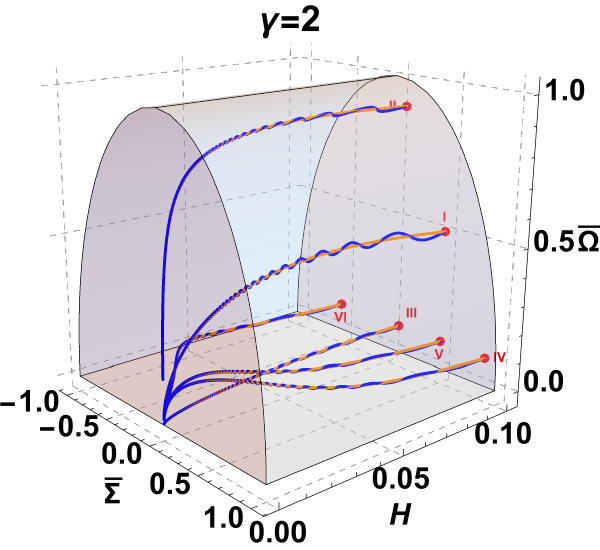}
    \caption{Some numerical solutions for full system \eqref{fullsyst-a}-\eqref{fullsyst-f} (blue) and time averaged system \eqref{BV} (orange) for different values of $\gamma.$ The initial conditions for each solution are given in Table \ref{TBV-2}}
    \label{FIG:2}
\end{figure}

\begin{figure}
    \centering
    \includegraphics[scale=0.6]{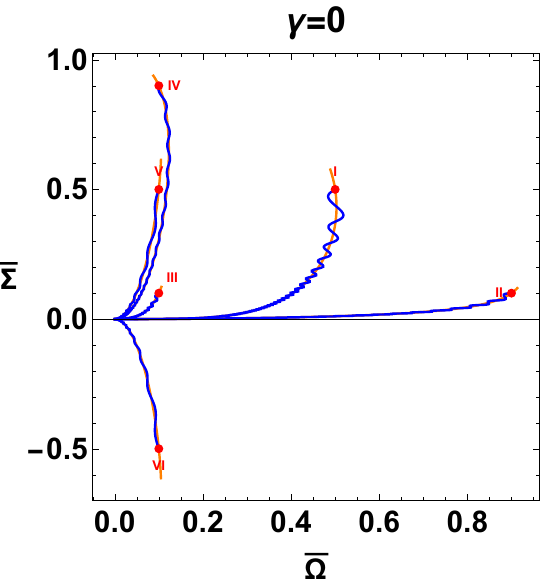}
    \includegraphics[scale=0.6]{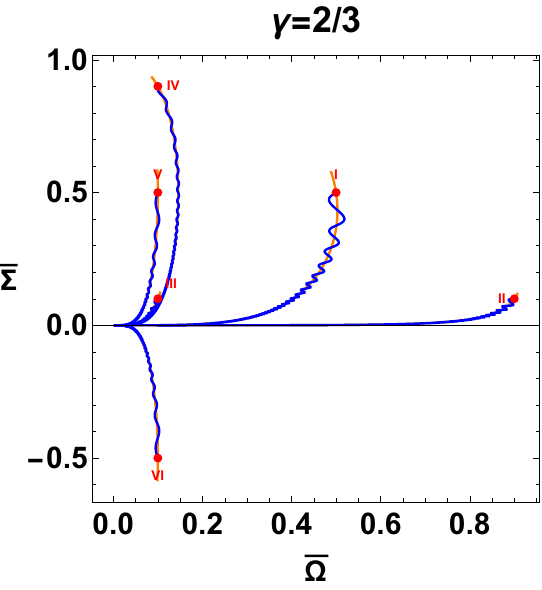}
    \includegraphics[scale=0.6]{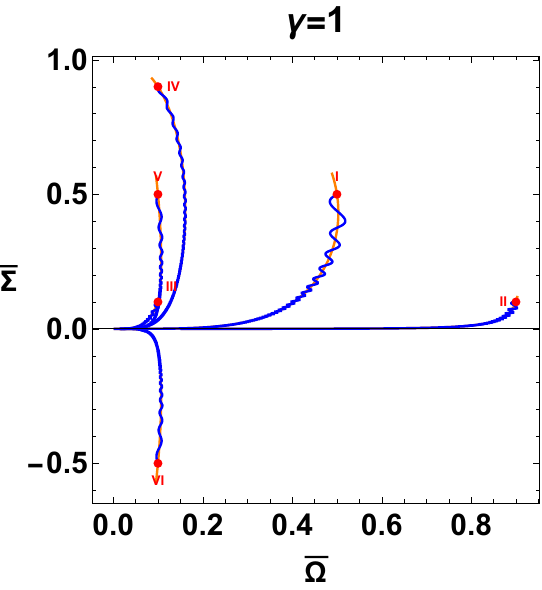}
    \includegraphics[scale=0.6]{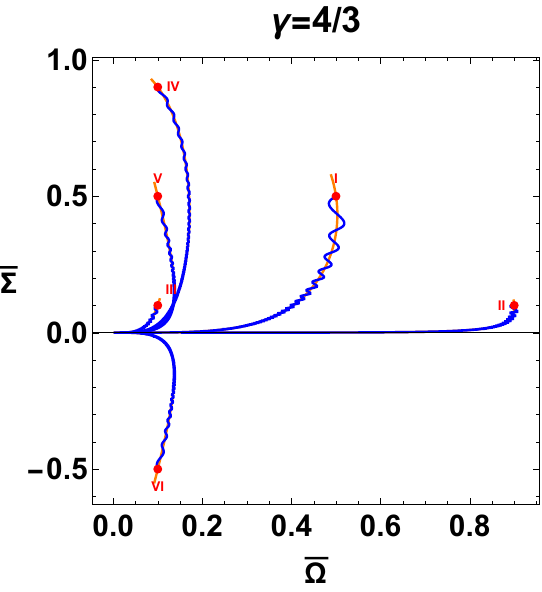}
    \includegraphics[scale=0.6]{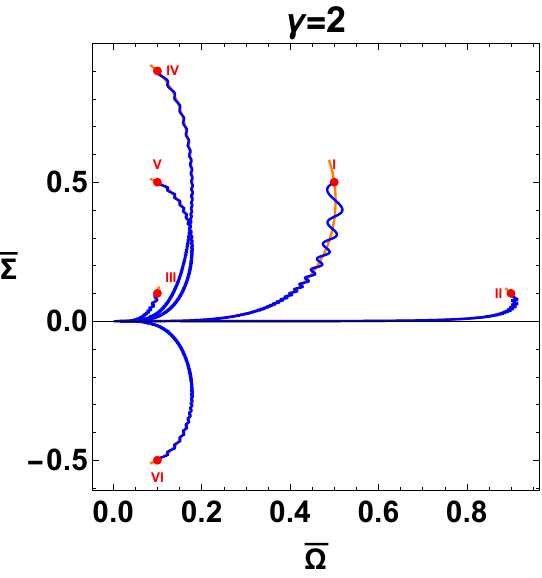}
    \caption{2 dimensional projection in the $(\overline{\Omega}, \overline{\Sigma})$ plane for different values of $\gamma,$ with the same set of initial conditions given in Table \ref{TBV-2}}
    \label{FIG:3}
\end{figure}

Finally, we perform numerical integration for the full system \eqref{fullsyst-a}-\eqref{fullsyst-f} and time-averaged system \eqref{BV}. We use six different initial conditions to generate solutions; they are presented in Table \ref{TBV-2}. 
In Figure \ref{FIG:2}, we present the behaviour of the solutions starting in the initial conditions in a three-dimensional projection space $(\overline{\Omega},\overline{\Sigma}, H)$ for different values of the parameter $\gamma.$ We chose this space, including $H$, to see that it tends to zero or a constant value near zero. In these plots, it is clear that the solutions of the full system \eqref{fullsyst-a}-\eqref{fullsyst-f} (in blue) have an initial oscillatory behaviour but then behave similarly to the solutions of the time-averaged system \eqref{BV} (in orange). This suggests that the solutions have the same behaviour as $t\rightarrow \infty.$ In Figure \ref{FIG:3} we show some 2 dimensional projections in the $(\overline{\Omega},\overline{\Sigma})$ plane for the same values of $\gamma$ where we see the same behaviour as before. We will show that the solutions have the same limit as $t\rightarrow \infty.$
   \begin{table}[t]
\begin{center}
\begin{tabular}{|c|c|c|c|c|c|}
\hline
Sol. & $H(0)$ &$\overline{\Omega}(0)$ &  $\overline{\Omega}_k(0)$  & $\overline{\Sigma}(0)$ & $\phi(0)$\\ \hline 
$I$ & $0.1$ & $0.5$ & $0.5$ &$0.5$ &$0$  \\\hline 
$II$ & $0.1$ & $0.9$ & $0.1$ &$0.1$ &$0$  \\\hline 
$III$ & $0.1$ & $0.1$ & $0.9$ &$0.1$ &$0$  \\\hline 
$IV$ & $0.1$ & $0.1$ & $0.1$ &$0.9$ &$0$  \\\hline 
$V$ & $0.1$ & $0.1$ & $0.1$ &$0.5$ &$0$  \\\hline 
$VI$ & $0.1$ & $0.1$ & $0.1$ &$-0.5$ & $0$ \\\hline
\end{tabular}
\end{center}\caption{Set of initial conditions for the numerical solutions of systems \eqref{fullsyst-a}-\eqref{fullsyst-f} and \eqref{BV} presented in Fig \ref{FIG:2}.}
\label{TBV-2}
\end{table}
 
\section{Conclusions}
\label{conclusions}
Based on the results presented, it is evident that the oscillations resulting from harmonic functions can be effectively smoothed out, which simplifies the problem at hand. This approach is equally beneficial for linear cosmological perturbations. It is essential to note that the coefficients of equations governing linear cosmological perturbations include background quantities. Therefore, a thorough understanding of background dynamics is crucial for further perturbation analyses.

In section \ref{Sect:2}, we review the main results of perturbation theory and some applications to different cosmological models.

In section \ref{sec3}, we study the evolution of cosmological perturbations in a vacuum with a generic perturbed FLRW metric. Here we obtained averaged equations \eqref{2d-a}-\eqref{2d-b} with two equilibrium points
\begin{enumerate}
    \item the attractor $(\overline{\Omega},\overline{z})=(0,0)$ and
    \item the source $(\overline{\Omega},\overline{z})=(1,0).$
\end{enumerate}
We also find that the system \eqref{2d-a}-\eqref{2d-b} is integrable and that in the limit $\tau \rightarrow \infty$ both $\overline\Omega (\tau )\rightarrow 0$ and $\overline z(\tau ) \rightarrow 0$ meaning that we are in the presence of long wavelength perturbations. $H (\tau )\rightarrow e^{c_1} c_3, y (\tau ) \simeq c_4-\frac{2 e^{-2 c_1} \left(\omega ^2-1\right) (3 \tau +2
   c_1)}{3 c_3{}^2} \rightarrow \left(\omega ^2-1\right) (-\infty ),  \Phi (\tau )\rightarrow c_5-\frac{e^{c_1} c_3 \left(\omega ^2-1\right)^3}{4 \omega   ^3}.$

   Finally, in section \ref{Sect:3.4}, we study the stability analysis of the equilibrium points for the Bianchi V metric and the behaviour of the solutions of the complete and time-averaged systems.

   In section \ref{analysis}, we performed a detailed dynamical system analysis for the guiding system \eqref{guidingBV} and found that the late time attractors for the Bianchi V metric (of both the full and averaged systems) are
   \begin{enumerate}
    \item[a)] The flat matter-dominated FLRW solution $P_1$ for $0\leq \gamma <\frac{2}{3}.$
    \item[b)] The curvature-dominated Milne solution $P_3$ for 
 $\frac{2}{3}<\gamma\leq 2.$ 
 \end{enumerate}
That result is consistent with that the Bianchi V universe can isotropize to the FLRW with a negative spatial curvature.

   In section \ref{numerical}, we computed some numerical solutions for both the entire system \eqref{fullsyst-a}-\eqref{fullsyst-f} (blue) and time-averaged system \eqref{BV}. We showed that the solutions have the same behaviour as $t\rightarrow \infty$, which means that we can apply perturbation methods to study the dynamics of cosmological models, in particular the Bianchi V metric to review a more simple version of the dynamical system with solutions that behave in the same way without the initial oscillatory behaviour.
\begin{acknowledgments}

A. D. Millano was supported by ANID Subdirección de Capital Humano/Doctorado Nacional/año 2020 folio 21200837, Gastos operacionales proyecto de tesis/2022 folio 242220121, and Vicerrectoría de Investigación y Desarrollo Tecnológico (VRIDT) at Universidad Católica del Norte(UCN). 
G. L.  thanks to VRIDT-UCN by the scientific and financial support through Resolución VRIDT No. 026/2023 and Resolución VRIDT No. 027/2023, and the support of Núcleo de Investigación Geometría Diferencial y Aplicaciones, Resolución Vridt No. 096/2022.

\end{acknowledgments}

\end{document}